\documentclass[pdflatex,sn-nature]{sn-jnl}
\usepackage[T1]{fontenc}
\usepackage[utf8]{inputenc}
\usepackage{amsmath}
\usepackage{amssymb}
\usepackage{xspace}
\usepackage{booktabs}
\usepackage{graphicx}
\usepackage{xcolor}
\usepackage{pgfplots}
\definecolor{clrn}{HTML}{0072B2}\definecolor{csub}{HTML}{D55E00}\definecolor{cmv}{HTML}{009E73}
\usepackage{enumitem}
\pgfplotsset{compat=1.17, every axis plot post/.append style={every mark/.append style={solid}}}
\setlist{itemsep=2pt, topsep=4pt}
\setcitestyle{super}

\newcommand{\mev}{meV/atom}
\newcommand{\Nel}{N_\mathrm{el}}
\newcommand{\Npe}{\tilde N_\mathrm{el}}
\newcommand{\lin}{\mbox{lin}\xspace}
\newcommand{\net}{\mbox{net}\xspace}
\newcommand{\pltwo}{\mbox{pl2}\xspace}
\newcommand{\plthree}{\mbox{pl3}\xspace}
\newcommand{\mm}[2]{\mbox{#1-}#2}
\newcommand{\lrn}{\mbox{LRN}\xspace}
\newcommand{\flrn}{\mbox{FIX-LRN}\xspace}
\newcommand{\fsub}{\mbox{FIX-SUB}\xspace}
\newcommand{\fsubp}{\mbox{FIX-SUB$^{+}$}\xspace}
\newcommand{\fmv}{\mbox{FIX-MV}\xspace}
\newcommand{\fmvp}{\mbox{FIX-MV$^{+}$}\xspace}
\newcommand{\fpt}{\mbox{FIX-PT$^{+}$}\xspace}
\newcommand{\fchr}{\mbox{FIX-CHR$^{+}$}\xspace}
\newcommand{\fmie}{\mbox{FIX-MIE$^{+}$}\xspace}
\newcommand{\frnd}{\mbox{FIX-RND}\xspace}
\newcommand{\fid}{\mbox{FIX-ID}\xspace}
\providecommand{\doi}{}\renewcommand{\doi}[1]{\href{https://doi.org/#1}{doi:#1}}

\begin{document}

\title[How many dimensions does chemical identity need?]{How many dimensions does chemical identity need? Learned element spaces from structures and compositions}

\author*[1,2]{\fnm{G.-M.} \sur{Rignanese}}\email{gian-marco.rignanese@uclouvain.be}

\affil*[1]{\orgdiv{Institute of Condensed Matter and Nanosciences}, \orgname{UCLouvain}, \orgaddress{\city{Louvain-la-Neuve}, \country{Belgium}}}
\affil[2]{\orgname{WEL Research Institute}, \orgaddress{\city{Wavre}, \country{Belgium}}}

\abstract{Machine-learning models represent chemical elements as vectors, but how many dimensions this representation needs has not been measured across the periodic table.
We force all chemical information through a bottleneck of learned pseudoelements and measure held-out accuracy as its dimension grows, along two routes: from crystal structures, with an alchemically compressed atomic density, and from compositions alone.
On the Alexandria database, with whole chemical systems held out, both routes agree for 86 elements.
Four learned dimensions capture most of what formation energies need and eight to twelve capture nearly all of it, an optimum that does not move when the training set grows from $10^5$ to $10^6$ structures.
Spaces built from substitution statistics, and fifteen published element embeddings including those of universal interatomic potentials, need at least twice as many dimensions.
A four-dimensional space learned once can be reused: kept fixed, it outperforms the substitution-based space and the best published embeddings on small training sets, for other properties (energy above the hull, magnetization), and for other functionals (PBEsol, SCAN).
Its directions are an electronegativity scale, an atomic size, and the filling of the d shell, and it reproduces the element parameters of Miedema's model of alloy formation enthalpies.}

\maketitle

\section{Introduction}

Chemical similarity between elements has been encoded in many ways.
Examples include the Pettifor scale\cite{pettifor1984} and its optimized one-dimensional variants\cite{glawe2016}, substitution statistics mined from crystal databases\cite{hautier2011,wang2021}, non-orthogonal Pettifor embeddings\cite{cerqueira2026}, word embeddings\cite{tshitoyan2019}, and element vectors extracted from trained graph networks or composition models\cite{zhou2018,goodall2020,mattervial}.
The ElementEmbeddings package\cite{onwuli2023} collects many of these, from vectors of tabulated properties\cite{ward2016,oliynyk2016,kusaba2022,kusaba2023} and of the literature\cite{tshitoyan2019,weston2019,gupta2022} to those learned by models of crystal structures\cite{chen2019,antunes2022,jang2024,antunes2024,park2025} and by universal interatomic potentials\cite{deng2023,batatia2024,neumann2024,park2024}.
These representations range from one to several hundred dimensions, and how many dimensions chemical identity really needs remains open.
The question matters in practice.
The cost of element-resolved atomic descriptors grows with the square of the number of channels\cite{bartok2013,musil2021}, so the size of the element space limits how many elements a model can treat.
A low-dimensional space is also easier to interpret and to transfer between properties.

Alchemical compression\cite{willatt2018,lopanitsyna2023} gives a direct way to measure this dimension.
The element-resolved neighbor density is contracted by a matrix $U$ from $\Nel$ elements to $\Npe$ pseudoelements before any features are built.
The model then sees chemistry only through $\Npe$ channels, and $\Npe$ becomes a controllable dimension of the element space.
Willatt et al.\ found two to four pseudoelements sufficient for a dataset of elpasolites\cite{willatt2018}.
Lopanitsyna et al.\ found three to four for high-entropy alloys of 25 transition metals\cite{lopanitsyna2023}.
Both studies covered restricted chemical spaces.

Here, we measure the dimension across 86 elements on the Alexandria database\cite{schmidt2023}, with whole chemical systems held out from training.
We do this along two independent routes, one from structures and one from compositions.
Along both routes we compare learned element spaces with fixed ones derived from chemical knowledge and with fifteen published element embeddings.
We also identify two ways in which a flexible model can bypass the bottleneck, which would make the measured dimension meaningless, and control for both.

We find that four learned dimensions capture most of the chemical information that formation energies need, and that eight to twelve capture nearly all of it, with up to $10^6$ training structures.
No fixed or published element space comes close at equal dimension.
A learned four-dimensional space is also reusable across training-set sizes, properties, and exchange-correlation functionals, and its directions have a clear physical reading.

The properties studied describe materials at equilibrium: mainly the formation energy, with the energy above the hull, the magnetization, the band gap, and the volume as further tests.
Forces away from equilibrium, vibrational and response properties, and the distinction between oxidation states of one element are not probed.
A model that has to describe all of these needs an element space containing those of the individual properties, so the dimensions reported here are lower bounds for such a model.

\section{Results}

\subsection{Measuring the dimension of an element space}
\label{sec:measure}

\begin{figure}[htbp]
\centering
\includegraphics[width=\textwidth]{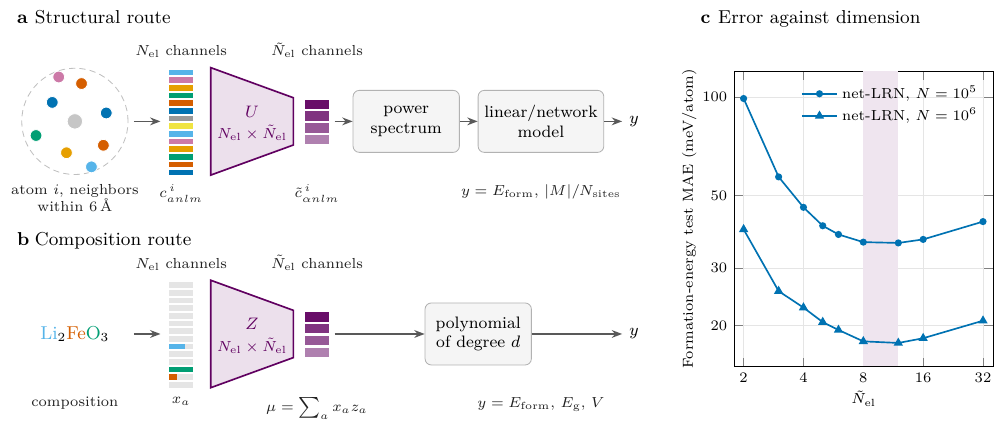}
\caption{Measuring the dimension of an element space.
(a) Structural route: the neighbor density of each atom, with one channel per element, is contracted by a matrix $U$ onto $\Npe$ channels before the power spectrum is built, so the model sees the neighbors' chemistry only through $\Npe$ channels.
(b) Composition route: the element fractions $x_a$, one channel per element, are contracted by a matrix $Z$, whose rows are the element vectors $z_a$, onto $\Npe$ channels $\mu$, and a polynomial of fixed degree in $\mu$ predicts the property.
In both routes, $U$ or $Z$ is learned or fixed from chemical knowledge, and $\Npe$ is varied.
The property $y$ is the formation energy, and also the magnetization per atom in the structural route and the band gap and volume per atom in the composition route.
(c) Formation-energy test MAE of the network model for 86 elements on held-out chemical systems, with a learned matrix (\lrn), trained on $N = 10^5$ and $10^6$ structures (20 epochs at $N = 10^6$; details in Fig.~\ref{fig:structure}; both axes logarithmic); four dimensions hold most of the chemical information, and the error is lowest at $\Npe = 8$--12 (shaded).}
\label{fig:overview}
\end{figure}

We measure the dimension along two routes that share no model and no input other than the elements (Fig.~\ref{fig:overview} and Methods).
In the structural route, the element-resolved neighbor density of each atom is contracted by an $\Nel \times \Npe$ matrix $U$ onto $\Npe$ pseudoelements, $\tilde c^{\,i}_{\alpha n l m} = \sum_{a} U_{a\alpha}\, c^i_{a n l m}$, before an invariant power spectrum is built. Two models then predict the formation energy from the power spectrum: a linear ridge model and a per-atom neural network.
In the composition route, each element receives a vector $z_a \in \mathbb{R}^{\Npe}$, and a polynomial of fixed degree in the composition-weighted mean $\mu = \sum_a x_a z_a$ predicts formation energy, band gap, and volume.
In both routes, $U$ or $Z$ is either learned or fixed from chemical knowledge.
The models depend only on the space spanned by the columns of $U$ or $Z$, not on the columns themselves, so individual pseudoelements are not defined; what is defined is an element space of dimension $\Npe$, in which each element has $\Npe$ coordinates.
Data come from the calculations of Alexandria\cite{schmidt2023} with the Perdew--Burke--Ernzerhof (PBE) functional\cite{perdew1996}, for two element sets, 25 d-block metals and 86 common elements (Methods).
Each chemical system is assigned to training, validation, or test by a hash of its element set, so that no test compound shares its element set with a training compound.
Errors are mean absolute errors (MAE) on the test set. The training-set size $N$ is the number of training structures in the structural route and the number of training compositions in the composition route.

\textbf{Element spaces compared.}
We name each matrix by how it is obtained (Methods).
\lrn denotes a matrix learned jointly with the model, and FIX a matrix fixed before training, followed by its source, as in \fsub.
A superscript $^{+}$ denotes a fixed matrix completed by a constant column, which carries the element-blind total density that centered components discard.
In tables and figures, the two structural models are abbreviated \lin and \net and the composition model \pltwo or \plthree for a polynomial of degree $d = 2$ or 3, and the model is prefixed to the matrix when it matters, as in \mm{lin}{\fsubp}, \mm{net}{\lrn}, or \mm{pl3}{\lrn}.
In the composition model, fixed matrices are used without the constant column, whose role its per-element term already plays.
The fixed matrices are:
\begin{itemize}
  \item \fsub and \fsubp, the leading principal components of the element substitution matrix of Wang et al.\cite{wang2021}, mined from the Inorganic Crystal Structure Database and used by Cerqueira et al.\ as non-orthogonal Pettifor embeddings\cite{cerqueira2026} (98 dimensions);
  \item \fmv and \fmvp, the same for the MatterVial element vectors, averages of the atomic features of ten pretrained graph-network models\cite{mattervial} (832 dimensions);
  \item the fifteen published embeddings of Table~\ref{tab:ee}, used with a constant column like \fsubp;
  \item \fpt, the position of each element in the periodic table;
  \item \fchr, three tabulated element characteristics (Pauling electronegativity, logarithm of the molar volume, and d-shell occupancy);
  \item \fmie, the parameters of Miedema's model\cite{miedema1980};
  \item \frnd, a random orthonormal matrix;
  \item \fid, the identity ($\Npe = \Nel$, no compression);
  \item \flrn, a matrix learned once with the model on $10^5$ PBE structures for the formation energy and then kept fixed for a new task.
\end{itemize}
As a floor, the composition baseline fits the target by ridge regression on the element fractions, without any element space; since every model includes this per-element term, the baseline is the limit $\Npe = 0$.

\textbf{Family purity.}
Besides accuracy, we ask how well a space groups the chemical families of textbook chemistry.
The family purity is the fraction of each element's five nearest neighbors, by Euclidean distance in the full space, that belong to its own family, averaged over the elements.
We use eight families: s-, d-, and f-block metals, p-block metals, metalloids, nonmetals, halogens, and noble gases (29 d-block and 21 f-block metals, but only four halogens and two noble gases among the 86 elements).
Random neighbors give 0.20, and a space that separated the families perfectly would give 0.96, because the halogens and noble gases have fewer than five other members.
The purity needs no property and no two-dimensional projection, so it tests whether a space agrees with chemical intuition independently of how it was obtained.
It does not test whether a space is useful for a property.
It ignores the order of elements within a family, which is what distinguishes, for instance, Ti from Ni for formation energies, and it says nothing about directions that no family label captures, such as an electronegativity scale that runs across families.
It is also coarse: separating the d- and f-block metals from the rest already earns much of the score, and the family labels are conventions.
We therefore use it as a check of chemical plausibility, next to the error, not as a measure of quality.

\subsection{Dimension from structures}

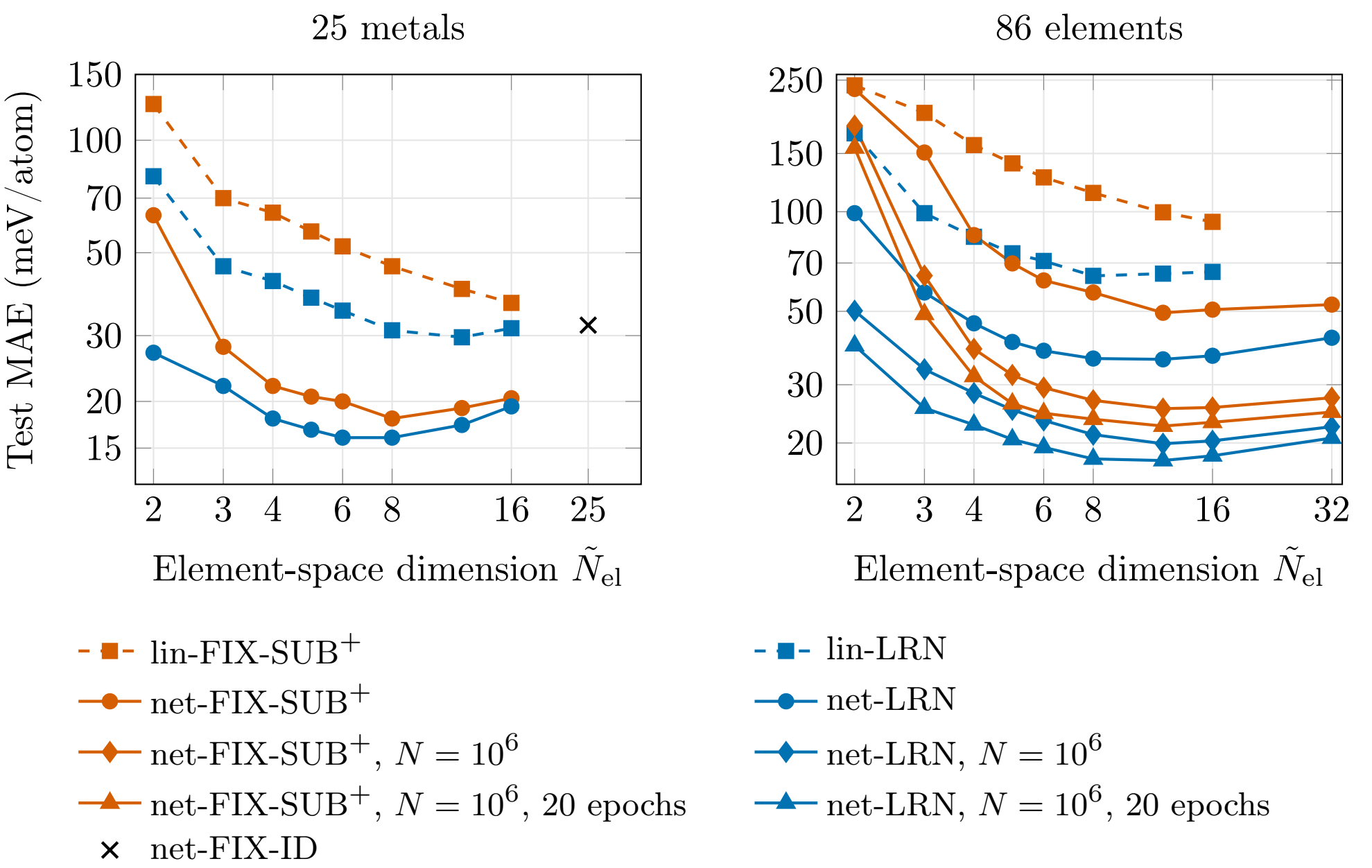
\begin{figure}[htbp]
\centering
\begin{tikzpicture}
\begin{axis}[width=0.49\textwidth, height=0.42\textwidth,
  xlabel={Element-space dimension $\Npe$}, xmode=log, xtick={2,3,4,6,8,16,25}, xticklabels={2,3,4,6,8,16,25},
  xmin=1.8, xmax=34, ymode=log, grid=major, grid style={gray!20},
  legend style={font=\scriptsize, draw=none, legend columns=2, /tikz/every even column/.append style={column sep=16pt}}, legend cell align=left,
  legend to name=structlegend,
  every axis plot/.append style={thick, mark size=1.8pt},
  title={25 metals}, ylabel={Test MAE (\mev)}, ymin=12, ymax=150,
  ytick={15,20,30,50,70,100,150}, yticklabels={15,20,30,50,70,100,150}]
  \addplot[csub, dashed, mark=square*, mark options={solid}] coordinates {(2,125) (3,70) (4,64) (5,57) (6,52) (8,46) (12,40.0) (16,36.7)};
  \addlegendentry{\mm{lin}{\fsubp}}
  \addplot[clrn, dashed, mark=square*, mark options={solid}] coordinates {(2,80.1) (3,46) (4,42) (5,37.9) (6,35) (8,31) (12,29.7) (16,31.4)};
  \addlegendentry{\mm{lin}{\lrn}}
  \addplot[csub, mark=*] coordinates {(2,63) (3,28) (4,22) (5,20.6) (6,20) (8,18) (12,19.2) (16,20.4)};
  \addlegendentry{\mm{net}{\fsubp}}
  \addplot[clrn, mark=*] coordinates {(2,27) (3,22) (4,18) (5,16.8) (6,16) (8,16) (12,17.3) (16,19.4)};
  \addlegendentry{\mm{net}{\lrn}}
  \addlegendimage{csub, thick, mark=diamond*, mark size=2.4pt}
  \addlegendentry{\mm{net}{\fsubp}, $N = 10^6$}
  \addlegendimage{clrn, thick, mark=diamond*, mark size=2.4pt}
  \addlegendentry{\mm{net}{\lrn}, $N = 10^6$}
  \addlegendimage{csub, thick, mark=triangle*, mark size=2.4pt}
  \addlegendentry{\mm{net}{\fsubp}, $N = 10^6$, 20 epochs}
  \addlegendimage{clrn, thick, mark=triangle*, mark size=2.4pt}
  \addlegendentry{\mm{net}{\lrn}, $N = 10^6$, 20 epochs}
  \addplot[black, only marks, mark=x, mark size=3pt] coordinates {(25,32)};
  \addlegendentry{\mm{net}{\fid}}
\end{axis}
\end{tikzpicture}\hfill
\begin{tikzpicture}
\begin{axis}[width=0.49\textwidth, height=0.42\textwidth,
  xlabel={Element-space dimension $\Npe$}, xmode=log, xtick={2,3,4,6,8,16,32}, xticklabels={2,3,4,6,8,16,32},
  xmin=1.8, xmax=34, ymode=log, grid=major, grid style={gray!20},
  every axis plot/.append style={thick, mark size=1.8pt},
  title={86 elements}, ymin=15, ymax=260,
  ytick={20,30,50,70,100,150,250}, yticklabels={20,30,50,70,100,150,250}]
  \addplot[csub, dashed, mark=square*, mark options={solid}] coordinates {(2,241) (3,199) (4,159) (5,140) (6,127) (8,114) (12,99.6) (16,93.2)};
  \addplot[clrn, dashed, mark=square*, mark options={solid}] coordinates {(2,172.6) (3,99) (4,84) (5,74.9) (6,71) (8,64) (12,65.0) (16,65.8)};
  \addplot[csub, mark=*] coordinates {(2,235) (3,151) (4,85) (5,69.8) (6,62) (8,57) (12,49.5) (16,50.6) (32,52.4)};
  \addplot[clrn, mark=*] coordinates {(2,99) (3,57) (4,46) (5,40.4) (6,38) (8,36) (12,35.8) (16,36.7) (32,41.6)};
  \addplot[clrn, mark=diamond*, mark size=2.4pt] coordinates {(2,50.2) (3,33.4) (4,28.3) (5,25.2) (6,23.4) (8,21.2) (12,19.9) (16,20.3) (32,22.4)};
  \addplot[csub, mark=diamond*, mark size=2.4pt] coordinates {(2,182) (3,64.1) (4,38.5) (5,32.1) (6,29.4) (8,26.9) (12,25.4) (16,25.6) (32,27.4)};
  \addplot[clrn, mark=triangle*, mark size=2.4pt] coordinates {(2,39.4) (3,25.5) (4,22.7) (5,20.5) (6,19.4) (8,17.9) (12,17.7) (16,18.3) (32,20.7)};
  \addplot[csub, mark=triangle*, mark size=2.4pt] coordinates {(2,156) (3,49.0) (4,31.8) (5,26.3) (6,24.6) (8,23.6) (12,22.5) (16,23.1) (32,24.8)};
\end{axis}
\end{tikzpicture}

\vspace{4pt}
\pgfplotslegendfromname{structlegend}
\caption{Formation-energy test MAE on held-out chemical systems against the dimension of the element space (logarithmic axes): color gives the matrix, line style the model (dashed, linear; solid, network), and symbol the training (squares and circles, $N = 10^5$; diamonds, $N = 10^6$ with 10 epochs; triangles, $N = 10^6$ with 20 epochs). Training uses $N = 10^5$ training structures, and $N = 10^6$ for the network model with 86 elements.
At $N = 10^5$, three seeds of the network model at $\Npe = 4$ and 8 have a standard deviation of at most 1.3~\mev; at $N = 10^6$, two seeds at $\Npe = 8$ and 12 agree to 0.1~\mev.
The \mm{lin}{\lrn} matrices are fit on 32,000 ($\Npe = 2$, 3, 5, 6, 12, 16) or 100,000 ($\Npe = 4, 8$) structures, one seed per point; a second seed at $\Npe = 3$ gives 48 (metals) and 91~\mev\ (86 elements).
The linear model stops at $\Npe = 16$: at $\Npe = 32$, its descriptor has 164,480 features per atom, and the ridge fit on $10^5$ structures would need about 270~GB of memory.
The network model's matrix starts from \fsubp.
The cross marks the network model with no compression (\fid).
Element-space names are defined in Section~\ref{sec:measure}.
A ridge model on the element fractions of the same structures (the composition baseline) gives 315 (metals) and 460~\mev\ (86 elements).}
\label{fig:structure}
\end{figure}

With a learned matrix, the error falls steeply up to $\Npe = 3$--4 and changes little beyond $\Npe = 6$ (Fig.~\ref{fig:structure}).
With the network model, going from $\Npe = 4$ to $\Npe = 8$ lowers the error by 11\% for the metals (18 to 16~\mev) and by 22\% for 86 elements (46 to 36~\mev).
The linear model has errors about twice as large and the same shape of curve.

Beyond $\Npe = 8$, adding dimensions no longer helps the network model, and the error rises past a minimum at $\Npe = 8$--12.
For 86 elements at $N = 10^5$, $\Npe = 8$, 12, 16, and 32 give 36, 36, 37, and 42~\mev.
For the metals, $\Npe = 8$, 12, and 16 give 16, 17, and 19~\mev.
Meanwhile the training error keeps falling (for 86 elements, from 21 to 10~\mev\ between $\Npe = 12$ and 32), so the extra channels are used to fit the training set rather than to generalize.
With ten times less data the effect is stronger: at $N = 10^4$, the error for 86 elements is 70~\mev\ at $\Npe = 8$ and 103~\mev\ at $\Npe = 32$.
The linear model, whose ridge penalty is chosen on validation data, does not rise but does not improve either: $\Npe = 8$, 12, and 16 give 64, 65, and 66~\mev\ for 86 elements and 31, 30, and 31~\mev\ for the metals, while the training error for 86 elements falls from 54 to 44~\mev\ between $\Npe = 8$ and 12.
The optimum near $\Npe = 8$ is therefore not specific to the network model.

A fixed matrix keeps improving over a longer range, because its leading directions are not the ones the energy uses.
In the linear model, \fsubp goes from 114 to 93~\mev\ for 86 elements and from 46 to 37~\mev\ for the metals between $\Npe = 8$ and 16.
In the network model, it flattens from $\Npe = 12$ (50, 51, and 52~\mev\ at $\Npe = 12$, 16, and 32 for 86 elements; 19 and 20~\mev\ at $\Npe = 12$ and 16 for the metals), above the learned matrix at $\Npe = 4$ (46 and 18~\mev).
Only in the linear model for the metals does \fsubp at $\Npe = 16$ beat the learned matrix at $\Npe = 4$ (37 against 42~\mev), and it remains above the learned matrix at $\Npe = 6$ (35~\mev).

Splitting the test error by compound class shows where the extra dimensions go (Table~S7).
For 86 elements, going from $\Npe = 4$ to 8 lowers the network model's error by 25\% for compounds of metals only, by 18\% for compounds of nonmetals with metals of a single common oxidation state (such as the alkali and alkaline-earth metals, Al, and Zn), and by 21\% for compounds of nonmetals with metals of several oxidation states (most d metals, Sn, Pb, and Ce); the linear model gives 27\%, 21\%, and 23\%.
The gain is thus not concentrated in compounds of multivalent elements, where a single vector per element must cover several oxidation states.
Compounds of metals only gain more within the 86 elements than the alloys of the 25 metals do (11\%): the extra dimensions serve the wider range of elements rather than a particular type of bonding.
Beyond $\Npe = 8$, the error of the metallic compounds stays flat (25, 25, and 25~\mev\ at $\Npe = 8$, 12, and 16), and the rise comes from the smaller classes, most of all from the 225 test compounds of nonmetals only (71 to 86~\mev).

More data lowers every error but barely moves the optimum.
With $10^6$ training structures for 86 elements, $\Npe = 2$, 3, 4, 8, 12, and 16 give 50, 33, 28, 21, 20, and 20~\mev, and $\Npe = 32$ gives 22~\mev\ (Fig.~\ref{fig:structure}).
Going from four to eight dimensions removes 25\% of the error, against 22\% at $N = 10^5$, and the curve is flat from 8 to 16 and rises again at 32, as at $N = 10^5$ (by 13\% from the minimum, against 16\%).
The dimension that formation energies need is therefore not an artifact of limited data: four dimensions remain a lower bound, and eight to twelve hold nearly all the chemical information the model can use.
The advantage of a learned matrix shrinks with data but persists at every dimension: \fsubp gives 182, 64, 39, 27, 25, 26, and 27~\mev\ at the same $\Npe$, so that the learned matrix is better by a factor of 1.4 at $\Npe = 4$ (against 1.8 at $N = 10^5$) and 1.3 at $\Npe = 8$ and 12 (against 1.6 and 1.4).
The substitution space has its optimum at the same place, $\Npe = 12$, and also rises at 32, but it needs twelve dimensions to reach the error that the learned space reaches with five (25~\mev).
These runs use 10 epochs; doubling the training to 20 epochs lowers the errors at $\Npe = 2$, 3, 4, 8, 12, 16, and 32 to 39.4, 25.5, 22.7, 17.9, 17.7, 18.3, and 20.7~\mev\ (\fsubp: 156, 49.0, 31.8, 23.6, 22.5, 23.1, and 24.8~\mev), but leaves the picture unchanged.
Going from four to eight dimensions still removes 21\% of the error, $\Npe = 8$ and 12 are equivalent, and the learned matrix keeps its advantage over \fsubp (a factor of 1.4 at $\Npe = 4$ and 1.3 at $\Npe = 8$); four learned dimensions now match twelve substitution ones (22.7 against 22.5~\mev).
A second seed at $\Npe = 8$ and 12 changes the error by 0.1~\mev.

The uncompressed network model for the metals (\fid, $\Npe = 25$) follows the same trend.
It reaches 32~\mev\ at $N = 10^5$ and 81~\mev\ at $N = 10^4$, against 16 and 34 at $\Npe = 8$.
Compression therefore also regularizes the model, and the uncompressed network model gives no ceiling on accuracy against which $\Npe = 8$ could fall short.
For 86 elements, an uncompressed descriptor (about $1.2 \times 10^6$ features per atom) cannot be computed.

The central-atom controls (Table~S1) show that the central atom's identity matters, but that the network model does not need more of it than $\Npe$ channels provide.
Removing the central atom from the density raises the error by 15--20\% for the metals and by about 40\% for 86 elements.
Giving the network model the central atom's full identity as a separate one-hot vector (one component per element) does not improve on the original model, even though it carries strictly more information.
These controls cannot tell whether the network model decodes the central atom's identity from its $\Npe$ channels or simply does not need more than $\Npe$ dimensions of it.
Either way, the network model's $\Npe$ bounds the chemistry of the neighbors, and the linear model and the composition route, which cannot exploit the central atom in this way, give the same dimension.

\subsection{Dimension from compositions}

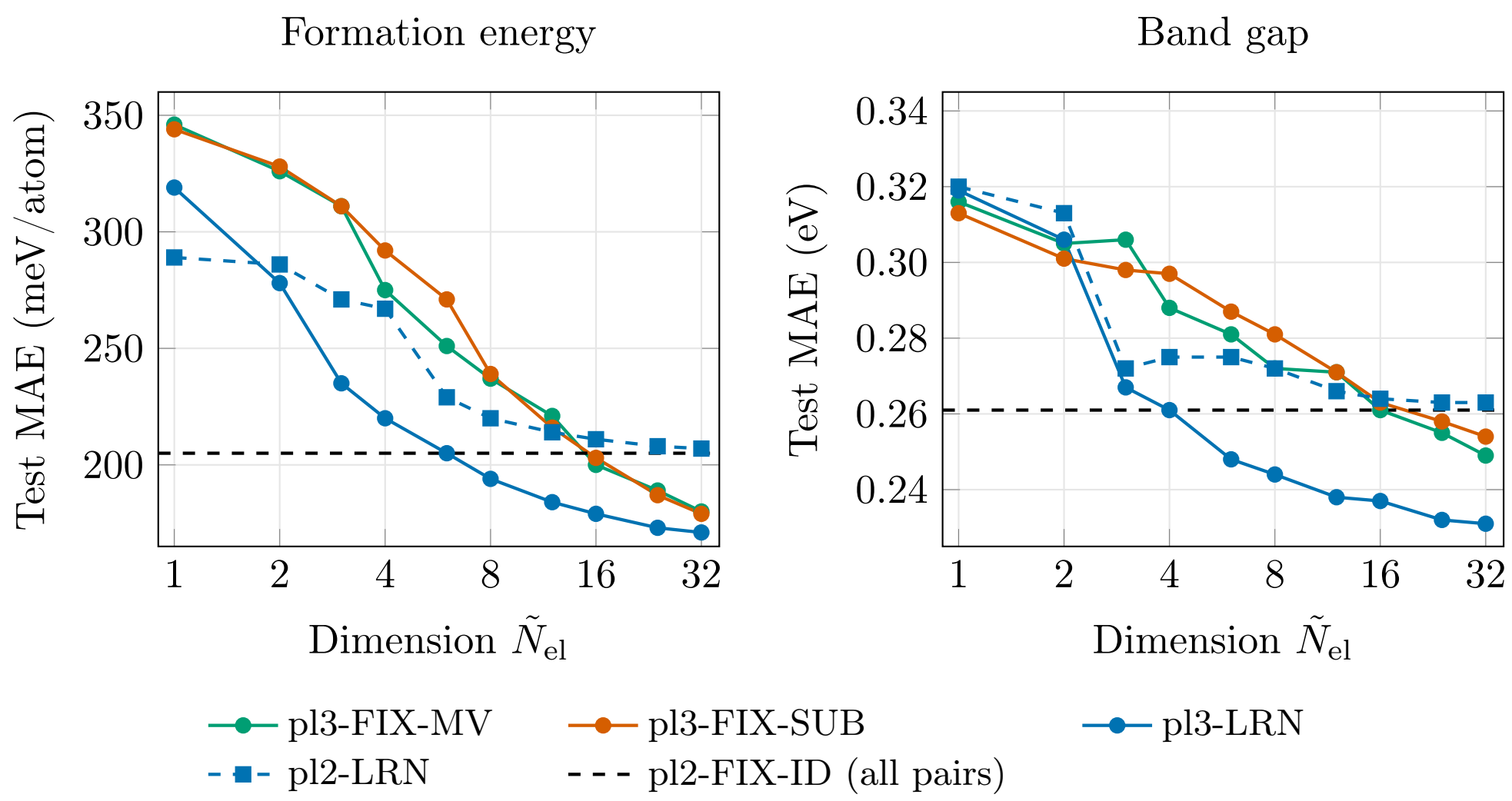
\begin{figure}[htbp]
\centering
\begin{tikzpicture}
\begin{axis}[width=0.49\textwidth, height=0.42\textwidth,
  xlabel={Dimension $\Npe$}, xmode=log, xtick={1,2,4,8,16,32}, xticklabels={1,2,4,8,16,32},
  xmin=0.9, xmax=36, grid=major, grid style={gray!20},
  every axis plot/.append style={thick, mark size=1.6pt},
  title={Formation energy}, ylabel={Test MAE (\mev)}, ymin=165, ymax=360]
  \addplot[cmv, mark=*] coordinates {(1,346) (2,326) (3,311) (4,275) (6,251) (8,237) (12,221) (16,200) (24,189) (32,180)};
  \addplot[csub, mark=*] coordinates {(1,344) (2,328) (3,311) (4,292) (6,271) (8,239) (12,216) (16,203) (24,187) (32,179)};
  \addplot[clrn, dashed, mark=square*, mark options={solid}] coordinates {(1,289) (2,286) (3,271) (4,267) (6,229) (8,220) (12,214) (16,211) (24,208) (32,207)};
  \addplot[clrn, mark=*] coordinates {(1,319) (2,278) (3,235) (4,220) (6,205) (8,194) (12,184) (16,179) (24,173) (32,171)};
  \addplot[black, dashed, domain=0.9:36, samples=2] {205};
\end{axis}
\end{tikzpicture}\hfill
\begin{tikzpicture}
\begin{axis}[width=0.49\textwidth, height=0.42\textwidth,
  xlabel={Dimension $\Npe$}, xmode=log, xtick={1,2,4,8,16,32}, xticklabels={1,2,4,8,16,32},
  xmin=0.9, xmax=36, grid=major, grid style={gray!20},
  every axis plot/.append style={thick, mark size=1.6pt},
  title={Band gap}, ylabel={Test MAE (eV)}, ymin=0.225, ymax=0.345,
  ytick={0.24,0.26,0.28,0.30,0.32,0.34},
  legend style={font=\scriptsize, draw=none, legend columns=3, /tikz/every even column/.append style={column sep=16pt}}, legend cell align=left,
  legend to name=complegend,
  yticklabel style={/pgf/number format/fixed, /pgf/number format/fixed zerofill, /pgf/number format/precision=2}]
  \addplot[cmv, mark=*] coordinates {(1,0.316) (2,0.305) (3,0.306) (4,0.288) (6,0.281) (8,0.272) (12,0.271) (16,0.261) (24,0.255) (32,0.249)};
  \addlegendentry{\mm{pl3}{\fmv}}
  \addplot[csub, mark=*] coordinates {(1,0.313) (2,0.301) (3,0.298) (4,0.297) (6,0.287) (8,0.281) (12,0.271) (16,0.263) (24,0.258) (32,0.254)};
  \addlegendentry{\mm{pl3}{\fsub}}
  \addplot[clrn, mark=*] coordinates {(1,0.319) (2,0.306) (3,0.267) (4,0.261) (6,0.248) (8,0.244) (12,0.238) (16,0.237) (24,0.232) (32,0.231)};
  \addlegendentry{\mm{pl3}{\lrn}}
  \addplot[clrn, dashed, mark=square*, mark options={solid}] coordinates {(1,0.320) (2,0.313) (3,0.272) (4,0.275) (6,0.275) (8,0.272) (12,0.266) (16,0.264) (24,0.263) (32,0.263)};
  \addlegendentry{\mm{pl2}{\lrn}}
  \addplot[black, dashed, domain=0.9:36, samples=2] {0.261};
  \addlegendentry{\mm{pl2}{\fid} (all pairs)}
\end{axis}
\end{tikzpicture}

\vspace{4pt}
\pgfplotslegendfromname{complegend}
\caption{Composition models (\pltwo and \plthree, polynomials of degree $d = 2$ and 3) for 86 elements, with element vectors learned jointly for formation energy, band gap, and volume (\lrn), compared with fixed principal components of the MatterVial (\fmv) and substitution (\fsub) embeddings weighted by composition, against the dimension of the element space (logarithmic axis).
Color gives the matrix and line style the model (dashed with squares, \pltwo; solid with circles, \plthree), as in Fig.~\ref{fig:structure}.
The uncompressed model with every pair interaction is \mm{pl2}{\fid}.
The composition baseline gives 428~\mev\ and 0.320~eV.
Test sets contain 324,382 held-out compositions.}
\label{fig:composition}
\end{figure}

The composition route gives an independent estimate that does not depend on the structural descriptor (Fig.~\ref{fig:composition}).
For 86 elements, the composition baseline gives 428~\mev, and the full pairwise model (\fid, $d = 2$) gives 205~\mev.
The 428~\mev\ differs from the 460~\mev\ of the structural data set because each composition appears here only once, as its lowest-energy polymorph (Methods).
With pair terms only, the learned curve flattens at $\Npe \approx 6$--8 (229 and 220~\mev).
With three-body terms, $\Npe = 6$ already equals the full pairwise model (205~\mev), $\Npe = 8$ gives 194~\mev, and the error keeps falling slowly, to 179, 173, and 171~\mev\ at $\Npe = 16$, 24, and 32.
The gain over the full pairwise model comes from the three-body terms, not from the compression itself: with pair terms only, no learned $\Npe$ beats it (207~\mev\ at $\Npe = 32$).
Compression makes the three-body terms affordable, with 56 cubic terms at $\Npe = 6$ against about $1.1 \times 10^5$ for 86 uncompressed elements.
The band gap and volume follow the same pattern: the learned $\Npe = 6$, $d = 3$ model reaches 0.248~eV and 1.62~\AA$^3$/atom, against 0.261 and 1.54 for the full pairwise model.
A second random start changes the errors by at most 2\%.

The fixed spaces need two to five times more dimensions.
With three-body terms, \fmv and \fsub need $\Npe = 16$ (200 and 203~\mev) to match the learned $\Npe = 6$, and $\Npe = 32$ (180 and 179~\mev) to match the learned $\Npe = 16$ (179~\mev), and stay above the learned $\Npe = 24$ and 32 (173 and 171~\mev).
For the band gap, \fmv at $\Npe = 32$ (0.249~eV) only matches the learned $\Npe = 6$ (0.248~eV), and \fsub stays above it (0.254~eV).
Both embeddings contain the relevant information, but their leading directions are not the ones these properties use.

For the 25 metals, the composition route carries little beyond the fractions.
Every learned model flattens by $\Npe = 2$--3 (234--240~\mev), which is no worse than the full pairwise model at 240~\mev.
For these alloys, the structure carries most of the chemical information that matters.

\textbf{The dimension depends on the task.}
Learning one property at a time needs fewer dimensions than learning all three together (Table~S2).
For formation energy alone, most of the gain is reached by $\Npe = 3$--4, while the band gap and volume keep improving up to $\Npe = 6$.
The joint model at $\Npe = 8$ is about as accurate as the single-property models at $\Npe = 4$--6.
An element space shared across properties therefore needs more dimensions than any one property, and the four to six dimensions found from structures for formation energy should be read as a lower bound for a general-purpose space.

The structural route gives the same picture for a property of a different kind, the magnetization per atom (Table~S5).
For 86 elements at $N = 10^5$, a learned matrix gives 59, 53, 51, 49, 50, and 51~m$\mu_\mathrm{B}$/atom at $\Npe = 2$, 3, 4, 6, 8, and 12, against 138 for the composition baseline and 63 and 58 for \fsubp at $\Npe = 4$ and 8.
Magnetization therefore needs about as many dimensions as formation energy, four to six, with a minimum at $\Npe = 6$.
For the 25 metals, the learned matrix gives 57, 45, 41, 41, 43, and 46~m$\mu_\mathrm{B}$/atom over the same range, against 214 for the composition baseline and 50 and 46 for \fsubp, with the same minimum at $\Npe = 4$--6.
The magnetization space is not the formation-energy space, however: their subspace overlap (the mean squared cosine of the principal angles between the two spaces, Methods) at $\Npe = 4$ is 0.74 (chance 0.05), against at least 0.99 between seeds.

\subsection{Learned versus chemical knowledge}

With enough data, a learned matrix beats every fixed one of the same dimension.
The advantage is a factor of 1.5--1.9 in the linear model and 1.2--2.5 in the network model (Table~\ref{tab:arms}).
Among the fixed matrices, adding the constant column lowers the error of \fsub by 5--15\% in the linear model, which makes \fsubp the fair fixed baseline.
In the network model it helps for the metals (by 5--19\% for \fsub and \fmv) but not for 86 elements, where the matrices without it do as well or up to 7\% better (\fsub, 173 against 185~\mev\ at $N = 10^4$ and 81 against 85 at $N = 10^5$; \fmv, 160 against 161 and 79 against 85); against \fsub, the advantage of the learned matrix there is 1.8--2.3 instead of 1.8--2.5.
\fsub and \fmv agree within 5\% on 86 elements, while \fmv is 10--25\% worse on the metals at small $\Npe$.
The constant column helps \fmv as well for 86 elements (154 against 175~\mev\ in the linear model at $N = 10^5$), so that \fmvp matches \fsubp; in the network model at $N = 10^4$, \fmvp gives 161 (86 elements) and 53~\mev\ (metals), against 185 and 48 for \fsubp, and at $N = 10^5$ the two agree (85 and 23~\mev, against 85 and 22).

In the linear model, chemical knowledge helps when data is scarce.
At $N = 10^3$ and $\Npe = 4$, a matrix learned on the small set itself does no better than \fsubp on the metals ($113 \pm 13$ against $108 \pm 2$~\mev, three seeds) and worse on 86 elements (395 against 264).
The network model, whose matrix starts from \fsubp, does not show this trade-off: at $N = 10^3$ it reaches 60 (metals) and 167~\mev\ (86 elements), against 91 and 293 for \fsubp kept fixed.
Section~\ref{sec:transfer} shows that a matrix learned once on a large data set does better still.
The network model does depend on a chemical starting point.
Starting $U$ from a random matrix instead of \fsubp gives 52 instead of 33~\mev\ (metals) and 119 instead of 75 (86 elements) at $N = 10^4$.
The starting point helps the network model's optimization, but the space it reaches is not an artifact of the substitution prior.
The linear model, which starts from a random matrix and has no prior, learns a space that overlaps with the network model's at 0.65 for 86 elements and 0.84 for the metals (chance 0.05 and 0.16), and its d-filling direction matches the network model's (Table~S3 and Section~\ref{sec:meaning}).

On Materials Project structures, the network model also beats the linear model for 86 elements (107 against 148~\mev\ at $\Npe = 4$, $N = 10^5$).
For the metals it does worse (about 240 against 130), on a set of only 2,599 structures, which may reflect overfitting to Alexandria-like structures.

\begin{table}[htbp]
\centering
\small
\caption{Formation-energy test MAE (\mev) at $\Npe = 4$ for learned and fixed element spaces, seed 0. The Baseline column is the composition baseline, a ridge regression on the element fractions without any element space ($\Npe = 0$). Table~\ref{tab:transfer} gives means over three seeds for some of the same settings, which differ from these by at most 2~\mev.}
\label{tab:arms}
\setlength{\tabcolsep}{4pt}
\begin{tabular}{llrrrrrrr}
\toprule
Elements & Model & $N$ & Baseline & \frnd & \fmv & \fsub & \fsubp & \lrn \\
\midrule
25 metals & \lin & $10^4$ & 315 & 190 & 97 & 87 & 74 & 48 \\
25 metals & \lin & $10^5$ & 315 & 167 & 82 & 74 & 64 & 42 \\
25 metals & \net & $10^4$ & 315 & 112 & 56 & 53 & 48 & 33 \\
25 metals & \net & $10^5$ & 315 & 52 & 28 & 27 & 22 & 18 \\
\midrule
86 elements & \lin & $10^4$ & 461 & 383 & 202 & 199 & 184 & 113 \\
86 elements & \lin & $10^5$ & 460 & 345 & 175 & 168 & 159 & 84 \\
86 elements & \net & $10^4$ & 461 & 330 & 160 & 173 & 185 & 75 \\
86 elements & \net & $10^5$ & 460 & 187 & 79 & 81 & 85 & 46 \\
\bottomrule
\end{tabular}
\end{table}

\subsection{Published element embeddings}
\label{sec:ee}

Table~\ref{tab:ee} compares the fifteen published embeddings with the learned spaces at equal dimension, for 86 elements.
None of them comes close.
At $\Npe = 4$, the best gives 148~\mev\ in the linear model (CGNF, against 84 learned), 163 in the network model (XenonPy, against 75), and 280 in the composition model (CGNF, against 220).
At $\Npe = 8$, the best linear errors (110~\mev, CGNF and ORB v2) remain 1.7 times the learned 64.
The best published embeddings are only as good as the substitution statistics: they improve on \fsubp by at most 12\%.
The same holds for the 25 metals, where the best embeddings give 63--64~\mev\ in the linear model (ORB v2 and CHGNet, against 64 for \fsubp and 42 learned) and 42 in the network model (CHGNet, against 48 and 33).
The position of each element in the periodic table, used as a fixed space in the same way (\fpt), is a weaker baseline still: 192~\mev\ in the linear model at $\Npe = 4$, between random (345) and \fsubp (159), and 98 against 70 for \fsubp and 46 learned for the 25 metals at $\Npe = 3$.
It thus recovers 59\% of the error reduction from a random space to the learned one, against 71\% for \fsubp.
At its largest dimension, $\Npe = 7$, it gives 143~\mev, against 114 for \fsubp and 64 learned at $\Npe = 8$.
Encoding the table as one-hot periods and columns does worse (253~\mev\ at $\Npe = 4$), because columns are categories, which compress poorly into a few dimensions.

\begin{table}[htbp]
\centering
\footnotesize
\setlength{\tabcolsep}{4pt}
\caption{Published element embeddings for 86 elements: dimension, principal components for 90\% of the variance (equal weights), family purity of five nearest neighbors, fraction of the network model's learned four-dimensional space captured by the top 4 and top 8 components (over 84 elements, without Kr and Xe; Methods), and formation-energy test error (\mev) with the embedding as a fixed bottleneck of dimension $\Npe$: linear model ($N = 10^5$), network ($N = 10^4$, the size affordable for all embeddings), and composition model ($d = 3$, three properties jointly). Each embedding is reduced to its top principal components without using the target (unsupervised), as for \fsub and \fmv, and used with a constant column in the structural models. The purity is defined in Section~\ref{sec:measure} and the captured fraction in Methods. Chance levels are 0.20 for purity and 0.05 and 0.09 for the captured fractions.}
\label{tab:ee}
\setlength{\tabcolsep}{3.5pt}
\begin{tabular}{lrrrrrrrrr}
\toprule
 & & & & \multicolumn{2}{c}{Captured} & \multicolumn{2}{c}{\lin} & \net & \plthree \\
\cmidrule(lr){5-6}\cmidrule(lr){7-8}
Embedding & Dim. & 90\% & Purity & top 4 & top 8 & $\Npe=4$ & $\Npe=8$ & $\Npe=4$ & $\Npe=4$ \\
\midrule
\multicolumn{10}{l}{\emph{Tabulated properties}} \\
Magpie & 22 & 8 & 0.74 & 0.68 & 0.76 & 172 & 126 & 184 & 291 \\
Oliynyk & 44 & 8 & 0.75 & 0.62 & 0.70 & 177 & 130 & 182 & 289 \\
XenonPy & 58 & 12 & 0.75 & 0.70 & 0.79 & 159 & 112 & 163 & 291 \\
\multicolumn{10}{l}{\emph{Literature}} \\
mat2vec & 200 & 37 & 0.65 & 0.36 & 0.59 & 216 & 153 & 225 & 327 \\
MatScholar & 200 & 37 & 0.65 & 0.38 & 0.62 & 216 & 151 & 218 & 326 \\
MatSciBERT & 768 & 63 & 0.38 & 0.17 & 0.38 & 287 & 217 & 320 & 410 \\
\multicolumn{10}{l}{\emph{Models of crystal structures}} \\
MEGNet & 16 & 4 & 0.61 & 0.45 & 0.54 & 191 & 147 & 169 & 297 \\
SkipAtom & 200 & 25 & 0.71 & 0.52 & 0.71 & 206 & 136 & 219 & 295 \\
CGNF & 90 & 23 & 0.74 & 0.65 & 0.79 & 148 & 110 & 166 & 280 \\
CrystaLLM & 512 & 41 & 0.70 & 0.50 & 0.69 & 177 & 141 & 193 & 284 \\
Chemeleon & 512 & 35 & 0.52 & 0.35 & 0.60 & 209 & 166 & 213 & 330 \\
\multicolumn{10}{l}{\emph{Universal interatomic potentials}} \\
CHGNet & 64 & 26 & 0.72 & 0.59 & 0.81 & 179 & 118 & 182 & 290 \\
MACE-MP-0 & 128 & 33 & 0.60 & 0.53 & 0.79 & 172 & 121 & 174 & 292 \\
ORB v2 & 256 & 1 & 0.48 & 0.61 & 0.82 & 195 & 110 & 193 & 284 \\
SevenNet & 128 & 4 & 0.66 & 0.47 & 0.65 & 181 & 143 & 207 & 312 \\
\multicolumn{10}{l}{\emph{Periodic table and tabulated characteristics}} \\
Position (\fpt)\textsuperscript{c} & 7 & 4 & 0.77 & 0.53 & 0.65 & 192 & 143 & 204 & 318 \\
Characteristics (\fchr)\textsuperscript{d} & 3 & 3 & 0.59 & 0.63\textsuperscript{d} & 0.63\textsuperscript{d} & 177 & -- & 181 & 308 \\
\midrule
\multicolumn{10}{l}{\emph{References}} \\
Substitution (\fsubp)\textsuperscript{a} & 98 & 37 & 0.73 & 0.63 & 0.77 & 159 & 114 & 185 & 292 \\
MatterVial (\fmvp)\textsuperscript{b} & 832 & 43 & 0.75 & 0.66 & 0.76 & 154 & 108 & 161 & 275 \\
\frnd\textsuperscript{e} & 4--8 & 4 & 0.20 & 0.05 & 0.09 & 345 & 287 & 330 & 422 \\
\lrn\textsuperscript{e} & 4--8 & 3 & 0.72 & 1 & 1 & 84 & 64 & 75 & 220 \\
\bottomrule
\end{tabular}
\par\vspace{2pt}{\footnotesize \textsuperscript{a}\,Without the constant column (\fsub) in the composition model. \textsuperscript{b}\,Without the constant column (\fmv) in the composition model; in the linear model, \fmv gives 175 and 123. \textsuperscript{c}\,Period, group, position in the f row, and block, standardized; its rank of 6 allows $\Npe \le 7$, and the $\Npe = 8$ entry is for $\Npe = 7$. \textsuperscript{d}\,Pauling electronegativity, logarithm of the molar volume, and $(n-1)$d occupancy, standardized, plus a constant column in the structural models; the captured fraction is the overlap of these four columns with the learned space (0.25 with random columns in place of the three characteristics), so the top-4 and top-8 values coincide, and the four columns allow $\Npe \le 4$. \textsuperscript{e}\,The dimension is that of the bottleneck; the 90\% count and the purity are for $\Npe = 4$.}
\end{table}

The ranking follows how much of the learned space an embedding contains.
Across the fifteen embeddings, the captured fraction and the error have Spearman correlations of $-0.85$ (linear, $\Npe = 4$, top 4), $-0.95$ (linear, $\Npe = 8$, top 8), $-0.73$ (network), and $-0.81$ (composition model).
The property-based embeddings and CGNF place most of the learned space in their top four components (0.62--0.70) and do best at $\Npe = 4$.
The literature embeddings capture the least and do worst, MatSciBERT approaching a random matrix.
The interatomic potentials are an intermediate case.
The top four components of ORB v2, CHGNet, and MACE-MP-0 capture only 0.53--0.61 of the learned space, but their top eight capture 0.79--0.82, the highest values together with CGNF and XenonPy (0.79).
At $\Npe = 8$, ORB v2, CHGNet, and MACE-MP-0 accordingly rise to the best group (110--121~\mev\ in the linear model).
These models learn the directions that matter for energies, but their leading variance is spent elsewhere.

Neither the size of an embedding nor its family purity is a good guide.
Across the fifteen embeddings, the Spearman correlation between purity and error is only $-0.68$ (linear, $\Npe = 4$), $-0.50$ (linear, $\Npe = 8$), $-0.51$ (network), and $-0.55$ (composition model), weaker than for the captured fraction.
The position in the periodic table has the highest purity (0.77), because the families are defined by it, yet it is a weak space for energies (192~\mev\ in the linear model at $\Npe = 4$).
The number of components for 90\% of the variance ranges from 1 (ORB v2, from a single dominant direction) to 63 (MatSciBERT), and ORB v2, with a purity of 0.48, is among the most useful at $\Npe = 8$.
What matters is whether the few directions a property needs are among the leading ones, and the learned space provides the yardstick for this.

\textbf{Reconstruction dimension.}
A natural alternative estimate of the dimension is the size of the latent space needed to reconstruct an existing embedding, for instance with an autoencoder.
Table~S4 gives the number of principal components that explain 90\% of the variance, with every element weighted equally and with elements weighted by how often they occur in Alexandria compositions.
The second weighting gives the latent dimension of a linear autoencoder of composition vectors.

Reconstructing these embeddings takes 29--43 dimensions for 86 elements, five to ten times the number properties need.
The reconstruction dimension measures how much the representation holds, including variation that no property tested here uses.
It does not measure the dimension of chemical identity as it matters for properties.
Element vectors taken from a single MatterVial property model, by contrast, are low-dimensional (6--7 components for 90\% of the variance), consistent with the dimension found here.

\subsection{Reusing a learned element space}
\label{sec:transfer}

A learned space is useful in practice only if it can be learned once and reused.
We test this with \flrn, a four-dimensional matrix learned on $10^5$ PBE structures for the formation energy and kept fixed while training on smaller sets ($10^3$ and $10^4$ structures), on other properties (the energy above the convex hull and the magnetization per atom), or on formation energies from other exchange-correlation functionals, PBEsol\cite{perdew2008} and SCAN\cite{sun2015} (Table~\ref{tab:transfer}).
Alexandria provides PBEsol and SCAN calculations for about 400,000 of its compounds, which we split with the same chemical-system hash, so that test systems remain unseen by every model, including the one that learned the transferred matrix.
Each model reuses its own matrix (linear or network), and every entry is averaged over three seeds.

\begin{table}[htbp]
\centering
\footnotesize
\setlength{\tabcolsep}{4pt}
\caption{Reusing a four-dimensional element space. Test MAE with a matrix fixed from chemical knowledge (\fsubp), learned on the training set itself (\lrn), or learned once on $10^5$ PBE structures for the formation energy and then kept fixed (\flrn), for $N = 10^3$ and $10^4$ training structures.
Errors are in \mev, except for the magnetization (m$\mu_\mathrm{B}$/atom). Mean over three seeds; the standard deviation is at most 10 at $N = 10^3$ (13 for learned matrices in the linear model) and at most 4 at $N = 10^4$.
The best entry of each triplet is in bold.}
\label{tab:transfer}
\setlength{\tabcolsep}{3pt}
\begin{tabular}{lllrrrrrr}
\toprule
 & & & \multicolumn{3}{c}{$N = 10^3$} & \multicolumn{3}{c}{$N = 10^4$} \\
\cmidrule(lr){4-6}\cmidrule(lr){7-9}
Target & Model & Elements & \fsubp & \lrn & \flrn & \fsubp & \lrn & \flrn \\
\midrule
PBE $E_\mathrm{form}$ & \lin & 25 metals & 108 & 113 & \textbf{78} & 75 & \textbf{50} & \textbf{50} \\
 & & 86 elements & 264 & 395 & \textbf{178} & 185 & 114 & \textbf{101} \\
 & \net & 25 metals & 91 & \textbf{60} & 64 & 48 & 33 & \textbf{32} \\
 & & 86 elements & 293 & \textbf{167} & 195 & 184 & \textbf{74} & 75 \\
\midrule
PBEsol $E_\mathrm{form}$ & \lin & 25 metals & 104 & 115 & \textbf{75} & 67 & 46 & \textbf{43} \\
 & & 86 elements & 279 & 407 & \textbf{189} & 197 & 117 & \textbf{106} \\
 & \net & 25 metals & 96 & \textbf{59} & 78 & 47 & \textbf{31} & 35 \\
 & & 86 elements & 347 & \textbf{182} & 245 & 210 & \textbf{76} & 107 \\
\midrule
SCAN $E_\mathrm{form}$ & \lin & 25 metals & 126 & 129 & \textbf{91} & 87 & \textbf{55} & 58 \\
 & & 86 elements & 311 & 431 & \textbf{205} & 222 & 130 & \textbf{124} \\
 & \net & 25 metals & 108 & \textbf{76} & 87 & 56 & \textbf{40} & 42 \\
 & & 86 elements & 358 & \textbf{192} & 267 & 226 & \textbf{89} & 125 \\
\midrule
Energy above hull & \net & 25 metals & 85 & \textbf{68} & \textbf{68} & 48 & 39 & \textbf{38} \\
 & & 86 elements & 227 & \textbf{158} & 175 & 142 & \textbf{83} & 88 \\
Magnetization & \net & 25 metals & 136 & \textbf{110} & 116 & 81 & \textbf{67} & \textbf{67} \\
 & & 86 elements & 154 & \textbf{123} & 143 & 109 & \textbf{71} & 79 \\
\bottomrule
\end{tabular}
\end{table}

\flrn is never worse than \fsubp.
Within PBE, it beats \fsubp by factors of 1.3--2.5 at $\Npe = 4$ (by 4--33\% at $\Npe = 8$ and $N = 10^3$).
In the linear model it is also the best choice overall, because learning a matrix on $10^3$ structures overfits (178 against 395~\mev\ for 86 elements).
In the network model, which already starts from \fsubp and adapts the matrix gently, the transferred and the freshly learned matrices are within a few percent at $N = 10^4$, and adapting the matrix to the small set does better at $N = 10^3$ for 86 elements (167 against 195~\mev).
At $\Npe = 8$ and $N = 10^4$, the transferred matrix gives 45 and 95~\mev\ in the linear model (against 48 and 118 learned on the set) and 33 and 68 in the network model (against 34 and 70).

Published embeddings do not close the gap.
We repeated every entry of Table~\ref{tab:transfer} with three embeddings kept fixed at $\Npe = 4$ as in Section~\ref{sec:ee}: XenonPy and CGNF, the best fixed matrices there, and MACE-MP-0, from a universal interatomic potential.
\flrn beats all three in every entry, for both models and for every target and functional (Table~S8).
For 86 elements at $N = 10^4$ in the network model, the best of them gives 164~\mev, against 75 for \flrn and 184 for \fsubp.
The closest case is the magnetization of 86 elements with $10^3$ structures (148 against 143~m$\mu_\mathrm{B}$/atom), a difference within the spread between seeds.

The space also transfers to properties it was not trained on.
For the energy above the hull, the formation-energy matrix does as well as a matrix learned on that target (38 against 39~\mev\ for the metals, 88 against 83 for 86 elements), and 1.3--1.6 times better than \fsubp; the composition baseline gives 265 and 297~\mev.
For the magnetization, a property of a different physical origin whose own learned space differs from the formation-energy one (overlap 0.74), the formation-energy matrix equals the magnetization matrix for the metals (67~m$\mu_\mathrm{B}$/atom for both) and is within 12\% of it for 86 elements (79 against 71), against 81 and 109 for \fsubp and 215 and 139 for the composition baseline.
With $10^3$ structures, the transferred matrix still beats \fsubp for both properties (above the hull, 68 against 85~\mev\ for the metals and 175 against 227 for 86 elements; magnetization, 116 against 136 and 143 against 154~m$\mu_\mathrm{B}$/atom), but, as for the formation energy, a network matrix adapted on the small set does as well or better (68, 158, 110, and 123).

Across functionals, the PBE matrix still beats \fsubp in every case, by factors of 1.2--1.4 at $N = 10^3$ and 1.3--2.0 at $N = 10^4$ in the network model, and 1.4--1.9 in the linear model.
In the linear model it remains the best choice, or within 6\% of the best.
In the network model, however, a matrix adapted to the new functional now does clearly better: for 86 elements at $N = 10^4$, 76 against 107~\mev\ for PBEsol and 89 against 125~\mev\ for SCAN.
Part of the PBE element space is therefore specific to the functional, or to the PBE geometries it was learned on, and a network model gains by adapting it.
Starting the network model from the transferred matrix and letting it adapt does best of all: for 86 elements, it gives 158 and 176~\mev\ at $N = 10^3$ (PBEsol and SCAN), against 182 and 192 when it starts from \fsubp, and 72 and 86 against 76 and 89 at $N = 10^4$, with gains of 4--9\% for the metals (Table~S9).
A learned matrix is therefore best reused as the starting point of a model on a new functional, rather than kept fixed.

The composition route shows the same pattern.
With element vectors learned on all compositions and then fixed, $10^3$ training compositions give 239~\mev, 0.315~eV, and 1.87~\AA$^3$/atom at $\Npe = 4$.
This is against 294--312~\mev\ for \fmv and \fsub and 350~\mev\ for vectors learned on the small set.
The advantage persists at $10^4$ compositions (222 against 237--293~\mev) and disappears at $10^5$, where vectors learned on the set reach the same error.

\subsection{The learned element space}

Different models and starting points learn largely the same four-dimensional space, well above chance (Table~S3).
Two network runs with different seeds agree almost perfectly (overlap 0.99), but they started from the same \fsubp matrix, so this is not independent evidence.
The two routes also learn related spaces, although they share no model and no input other than the elements.
The eight-dimensional spaces learned from structures and from compositions have a subspace overlap of 0.54 (chance 0.10), a distance Spearman correlation of 0.56, and 59\% shared nearest neighbors (6\% for random spaces).

Nearest neighbors in the network model's four-dimensional space are mostly chemical.
For 86 elements, examples are Na (Li, K, Cd, Mg), Ca (Eu, Sr, Pr, Nd), O (N, S, F, H), S (Se, P, As, N), Si (Ge, P, As, Sn), and La (Pr, Nd, Pm, Sm).
For the metals, examples are Fe (Mn, Co, Cr, Ni), Y (Sc, Lu, Zr, Hf), W (Mo, Ta, Nb, Ru), and Ag (Au, Pd, Cu, Pt).
A few pairs are unexpected, such as Fe--Be and Na--Cd.
This is an energetic similarity.
It overlaps with substitution statistics without being the same, echoing the distinction Lopanitsyna et al.\ draw between thermodynamic and potential-based similarity\cite{lopanitsyna2023}.

\subsection{Maps of the element spaces}

\begin{figure}[htbp]
\centering
\includegraphics[width=\textwidth]{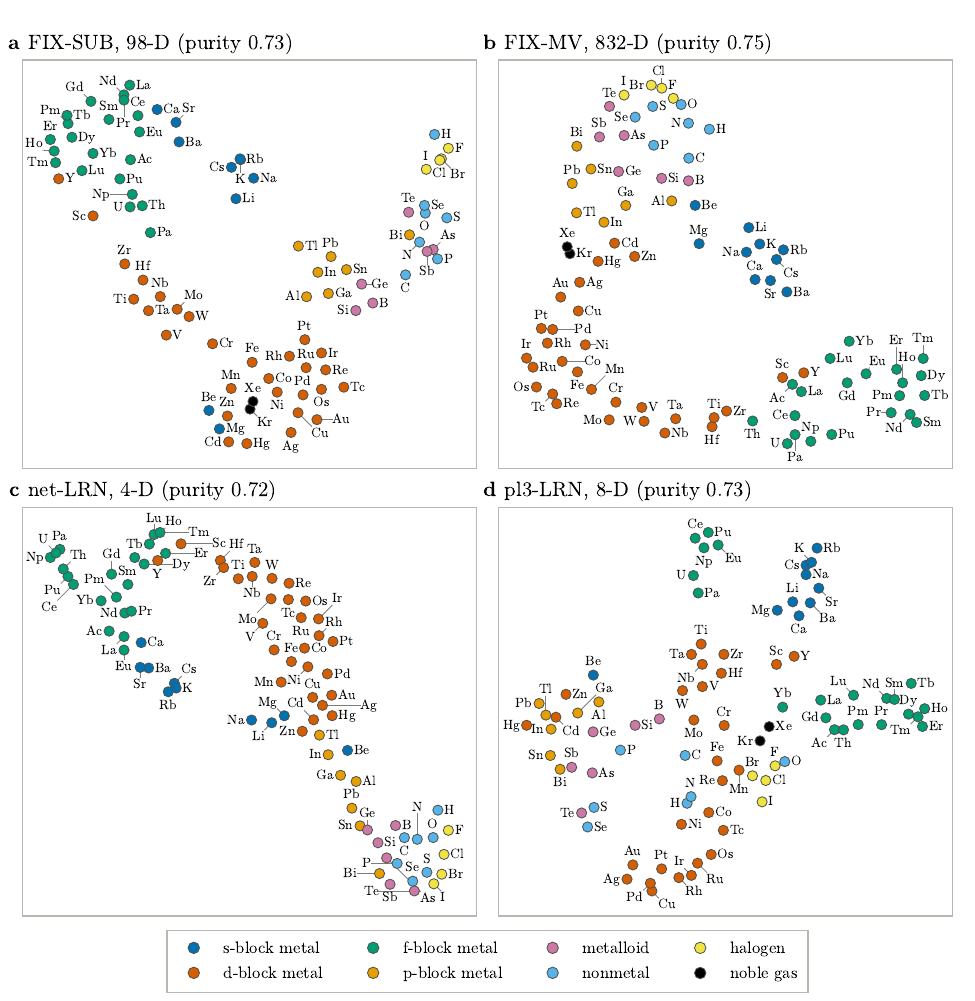}
\caption{Two-dimensional UMAP maps of 86 elements in (a) the substitution-based embedding \fsub\cite{wang2021,cerqueira2026}, (b) the MatterVial Global embedding \fmv\cite{mattervial}, (c) the four-dimensional \lrn space of the network model ($N = 10^5$), and (d) the eight-dimensional \lrn space learned from compositions ($d = 3$, three properties jointly).
The full embeddings are shown, without the constant column of \fsubp and \fmvp.
Learned matrices are orthonormalized before distances are computed.
The purity is the fraction of five nearest neighbors in the full space that share the element's family (0.20 for random neighbors).
Kr and Xe, which the network model sees in no or almost no training structure, are left out of map (c) but not of its purity.
Map axes have no physical meaning.}
\label{fig:maps}
\end{figure}

Following Cerqueira et al.\cite{cerqueira2026}, Fig.~\ref{fig:maps} projects element spaces to two dimensions with UMAP\cite{mcinnes2018} (10 neighbors, minimum distance 0.3, the same settings for every panel).
Panel~(a) uses the same substitution-based embedding as their map.
Because a projection can distort distances, we judge each space by its family purity (Section~\ref{sec:measure}), computed in the full space.

The learned spaces recover the chemical families as well as the much larger embeddings do.
The purity is 0.72 for the four-dimensional space learned from structures and 0.73 for the eight-dimensional space learned from compositions, against 0.73 for the 98-dimensional substitution embedding and 0.75 for the 832-dimensional MatterVial embedding.
All four maps separate the f-block, the s-block metals, the d-block metals, and the electronegative p-block elements.
The two learned maps also arrange the d-block metals along their rows and groups, for instance Ti, Zr, and Hf next to V, Nb, and Ta, and the noble metals Ru, Rh, Os, Ir, Pd, and Pt together.
The spaces differ in which families they keep whole.
MatterVial and the composition space keep the s-block metals together (0.88 of their five nearest neighbors are s-block metals), which makes map (b) look the cleanest, although MatterVial's purity over all families (0.75) only equals that of the best published embeddings (Table~\ref{tab:ee}).
The substitution and structural spaces split the s block (0.50--0.54): they place Ca, Sr, and Ba next to Eu, Yb, and the light lanthanides, and Be and Mg next to Zn, Cd, Al, and the 3d metals, the elements they replace in crystals.
Both learned spaces also split the f block in two, along different lines.
From compositions, Ce, Eu, Yb, and the early actinides Pa, U, Np, and Pu, which take more than one oxidation state, separate from the trivalent lanthanides, Ac, and Th.
From structures, Ce and the early actinides Th to Pu, whose f electrons take part in bonding, separate from the other lanthanides, Eu and Yb included.
Kr and Xe are left out of map (c): Kr appears in no training structure and Xe in only four (Methods), so their positions in the learned structural space are not set by the data.

\subsection{What the four dimensions mean}
\label{sec:meaning}

The models depend only on the space spanned by the columns of $U$ (or $Z$), so the individual axes carry no meaning.
We ask instead which elemental properties are linear functions of the four-dimensional space, using the leave-one-out $R^2$ of a linear fit with an intercept (Table~S6).
Random four-dimensional spaces give $R^2 \approx -0.08$.
Tabulated properties are taken from pymatgen\cite{ong2013}, and Miedema's parameters from matminer\cite{ward2018}.

\begin{figure}[htbp]
\centering
\includegraphics[width=0.93\textwidth]{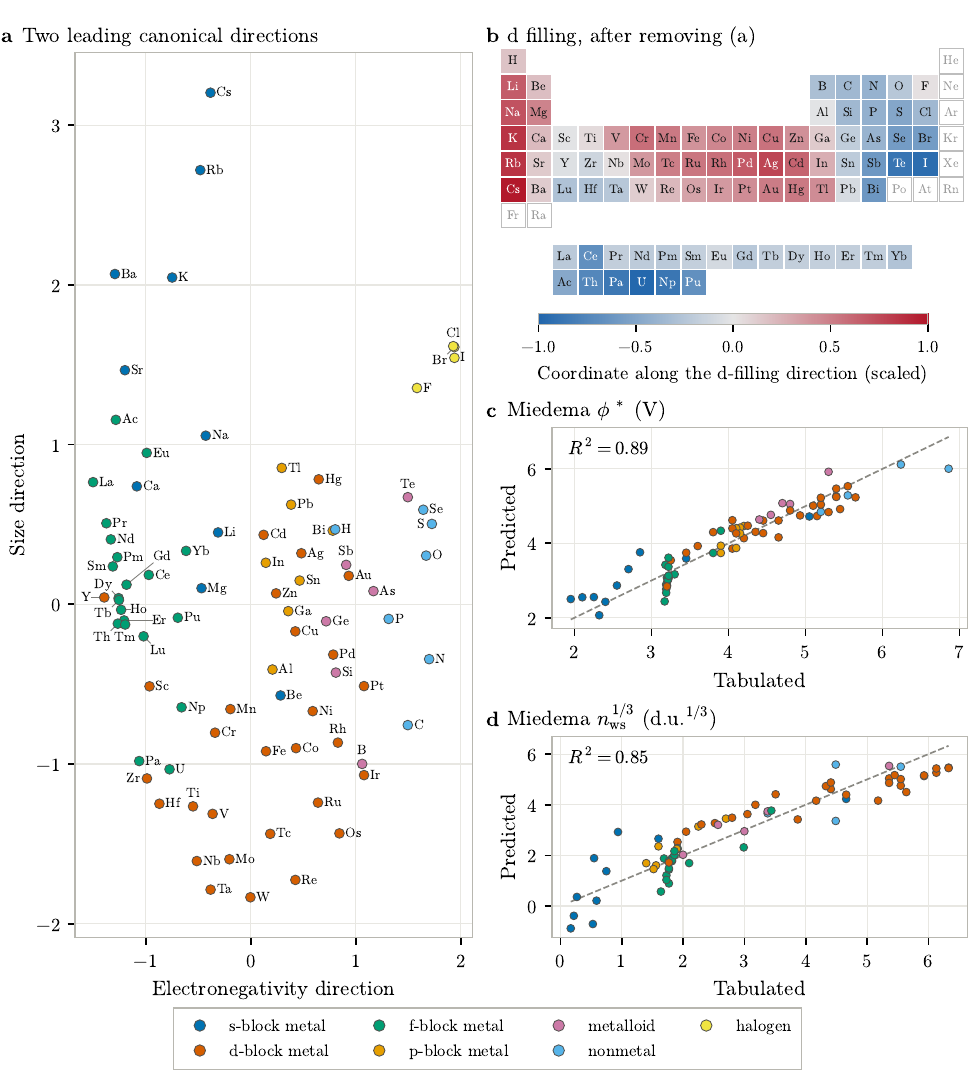}
\caption{Physical content of the learned structural spaces ($N = 10^5$).
Kr and Xe, whose positions are not set by the training data (Methods), are left out.
(a) Elements of the four-dimensional space projected on its two leading canonical directions against eight textbook properties, which correlate with electronegativity and with atomic size.
This plane holds 54\% of the spread of the four-dimensional space; principal components define no preferred plane, because the orthonormalized space has equal variance in its three non-constant directions.
(b) The four-dimensional space after projecting out these two directions: coordinate of each element along the remaining direction that best follows the d-electron count of the d-block metals ($r = 0.81$), drawn on the periodic table (scaled to the largest magnitude; empty cells are elements outside the 86).
(c, d) Miedema's parameters $\phi^*$ and $n_\mathrm{ws}^{1/3}$\cite{miedema1980} predicted linearly from the four-dimensional space (leave-one-out) against their tabulated values, for the 73 elements in Miedema's table.}
\label{fig:physics}
\end{figure}

The best-recovered quantities are the parameters of Miedema's semi-empirical model of alloy formation enthalpies\cite{miedema1980}.
These are an electronegativity-like chemical potential $\phi^*$, the electron density at the boundary of the Wigner--Seitz cell $n_\mathrm{ws}$, and the molar volume $V$.
All three learned spaces give $R^2 = 0.86$--0.94 for $\phi^*$ and 0.77--0.86 for $n_\mathrm{ws}^{1/3}$ (Fig.~\ref{fig:physics}c,d).
The textbook scales follow: Pauling electronegativity, Pettifor's Mendeleev number, the group number, ionization energy, and atomic radius reach 0.60--0.84, while the valence-electron count, period, and electron affinity stay at or below 0.5.

Canonical correlation analysis against eight textbook properties (84 elements with all of them) finds two directions shared by all three models.
The first has canonical correlations 0.95--0.97 and correlates with electronegativity ($r = 0.88$ in the network space) and the Mendeleev number ($r = 0.87$): it is a learned chemical scale.
The second has canonical correlations 0.89--0.91 and correlates with the molar volume ($r = 0.79$): it is an atomic size.
In this plane (Fig.~\ref{fig:physics}a), the large alkali and alkaline-earth metals sit at the top, the refractory 4d and 5d metals at the bottom, and the elements are ordered from electropositive to electronegative along the horizontal axis.

The other two directions are less clearly identified by canonical correlation.
Their canonical correlations, 0.72--0.81 and 0.52--0.63, remain above those of random spaces (0.24 and 0.15), but they do not match any single property in the set of eight.
After removing the directions that predict electronegativity, Mendeleev number, and molar volume, the remaining direction holds about a quarter of the spread (25--31\%).
Within the d block, it follows the filling of the d shell: its correlation with the number of d electrons over the 30 d-block elements is 0.83 in the network space, 0.70 in the linear space, and 0.78 in the composition space (the sign of a direction is arbitrary).
In the network space, it runs from the early actinides and Ce (U, Np, Pa, Th, Pu, Ce) to the late 4d and 5d metals (Ru, Rh, Pd, Ag, Os, Ir, Pt, Au), so the partly filled f shells continue the d-filling scale at its low end.
Accordingly, the four-dimensional spaces predict the d-electron count of the d-block elements with $R^2 = 0.81$--0.85 and the group number with $R^2 = 0.75$--0.82 (Table~S6).
Among free-atom quantities (Methods), they recover the LDA eigenvalues of the valence s and p levels with $R^2 = 0.69$--0.84, the highest occupied level with 0.51--0.59, and the calculated orbital radii with 0.66--0.82, but the energy of the d level of the d-block elements only with 0.36--0.53 (0.50--0.71 with eight dimensions), well below the d-electron count.
Outside the d block the models differ: the residual directions of the network and linear spaces agree ($r = 0.76$--0.78), while that of the composition space places the light covalent elements (B, C, N, O, H) at the end opposite the noble metals, and agrees less with the other two ($r = 0.20$--0.21).

This direction is not an accident of training.
Three network runs with different seeds learn the same space at $\Npe = 4$ and 8 (subspace overlap at least 0.99, for both element sets) and the same residual direction ($r \ge 0.99$), and each departs only partly from the \fsubp matrix they start from (overlap 0.81 with it at $\Npe = 4$ for 86 elements and 0.88 for the metals).
Because the seeds share this starting matrix, their agreement alone is not independent evidence.
The linear model, which starts from a random matrix, provides it: its residual direction agrees with the network model's at $r = 0.77$, against 0.09 for random directions.
For the 25 metals, where the full space already gives $R^2 = 0.81$ for the group number, the residual direction is less stable between seeds ($r = 0.80$--0.99) and matches no single property.

\textbf{What remains beyond electronegativity and size.}
Projecting the elements onto the complement of the two leading canonical directions removes what the space knows about electronegativity and period (leave-one-out $R^2 \le 0.21$) and most of what it knows about the group number ($R^2 = 0.20$ in the four-dimensional and 0.46 in the eight-dimensional network space, against 0.82 for both full spaces), and shows what the other dimensions encode.
In the four-dimensional network space, the remaining plane keeps 46\% of the spread, and its direction of d filling, found after removing the two canonical directions, has $r = 0.81$ over the d block (Fig.~\ref{fig:physics}b).
On the periodic table, this direction rises across the d block from Sc and Ti to Cu, Zn, and their heavier congeners, places the f block and the early actinides at its low end, and also separates the alkali metals from the p-block nonmetals.
In the eight-dimensional network space (Fig.~S2), the d filling is untouched by the projection ($R^2 = 0.92$ from the six remaining dimensions, as from all eight), and the columns of the periodic table remain together: 44\% of each element's five nearest neighbors share its column, against 48\% in the full space and 8\% for random vectors.
The d metals form an arc ordered by group, with Ti, Zr, and Hf, then V, Nb, and Ta, and so on to Zn, Cd, and Hg, while the alkali and alkaline-earth metals, the halogens, the pnictogens and chalcogens, and the lanthanides form separate clusters.
The space learned from compositions gives the same picture (40\% same-column neighbors, $R^2 = 0.77$ for the d filling from the residual).
Electronegativity and size thus order the elements across the periodic table, the d filling is a third continuous scale, and the remaining dimensions mostly keep each column together, as clusters rather than along a scale.

\textbf{The learned spaces and the periodic table.}
Fig.~S1 draws the four directions of the network space on the periodic table: electronegativity runs across it, size runs down and to the left, and the d filling runs across the d block.
On the size map, Eu and Yb stand apart from the other lanthanides, as expected for the two larger, divalent members of the series.
Along the rows and down the columns (Figs.~S3 and~S4), each direction is ordered within each block, but the order differs from block to block: down a column, the size grows for the alkali metals and the lanthanides give way to the actinides, while the p-block nonmetals move along the fourth direction.
Ce, which readily takes the 4+ state, leaves the lanthanides toward Th and the light actinides: the cosine between its displacement from the other lanthanides and that of Th is 0.87, and 0.80--0.93 in the eight-dimensional network space and in both composition spaces, though the linear model barely separates Ce from the other lanthanides.
Eu moves toward the alkaline-earth metals (cosine 0.94 with the displacement of Ca, Sr, and Ba, and 0.51--0.62 in the other network and composition spaces).
The nearest neighbors follow the table's topology.
Of each element's five nearest neighbors in the four-dimensional network space, 46\% share its column, 24\% its period, and 15\% are diagonal neighbors (one period and one group apart), against 8\%, 18\%, and 3\% for random vectors.
The column share is close to that of the substitution (51\%) and MatterVial (49\%) embeddings, and the diagonal share is higher than either (10\% and 13\%).
Several known irregularities of the table appear without being imposed.
The lanthanide contraction makes the 4d and 5d metals of a group closer than the 3d and 4d ones: the ratio of these distances, averaged over groups 4--12, is 0.63--0.88 in the learned spaces, against 1.07 for random ones.
Among diagonal pairs, Si is the nearest or second-nearest neighbor of B in every space, and Al the nearest of Be in the structural spaces.
For group 3, the structural spaces place Sc much closer to Lu than to La (Lu ranks 4th among the neighbors of Sc, La 27th to 51st), as the substitution embedding does, while the composition spaces lean the other way and Y sits between the two.
Hydrogen has no consistent place.
The learned space therefore keeps the topology of the periodic table but replaces its integer coordinates by physical scales, which matters for accuracy: the table's position alone recovers only 59\% of the gain of the learned space (Table~\ref{tab:ee}).
To separate wrong coordinates from missing information, we let the network model learn its element space only as a smooth function of the position in the table, a polynomial of degree 1, 2, or 3 in the table coordinates (7, 17, or 28 functions; Methods).
At $\Npe = 4$, this gives 68, 65, and 61~\mev, against 175, 119, and 94~\mev\ for random bases of the same size and 46~\mev\ for an unrestricted matrix; at $\Npe = 8$, degrees 2 and 3 give 51 and 42~\mev, against 77 and 64 for random bases and 36 unrestricted.
The table's coordinates therefore carry most of the information, in distorted form: a linear map of them already beats \fsubp (68 against 85~\mev\ at $\Npe = 4$), and smooth distortions close most of the remaining gap at $\Npe = 8$.
What no smooth function of the position supplies, 15--30\% of the error here, is consistent with the irregularities listed above, which an element's place in the table does not encode.

\textbf{Pseudoelements or element characteristics?}
Alchemical compression is usually described in terms of pseudoelements, mixtures of the real elements that the model sees instead of them\cite{willatt2018,lopanitsyna2023}.
Our results favor a different reading.
Because only the span of $U$ is defined, individual pseudoelements exist only after a basis has been chosen.
Even in the basis that makes them as localized as possible (varimax), the learned channels are as spread over the elements as those of random matrices: about 25 of 86 elements carry each channel at $\Npe = 4$ (25 for random matrices), 15 at $\Npe = 8$ (20), and 7 of 25 metals at $\Npe = 4$ (7).
They are signed contrasts between groups of elements, with one exception: a near-constant direction, which weights all elements equally and carries the element-blind total density.
The network model keeps it from its starting matrix, and the linear model, which starts from a random matrix, learns it on its own (share of the constant direction in the span 0.76, against 0.05 for random spaces).
The composition model does not need it, because its per-element linear term already plays this role, and its channels are as signed as random ones.
Read in the basis of the characteristics identified above, each contracted channel is a neighbor density weighted by an element characteristic: the plain density, a density weighted by electronegativity, by size, or by d filling.
This is the idea behind weighted atom-centered symmetry functions, in which element-dependent weights replace element channels\cite{gastegger2018}; here the weights are learned and their number is measured.
We therefore read $\Npe$ as the number of element characteristics that the property needs, rather than as a number of effective elements.

\textbf{Tabulated characteristics as a fixed space.}
If the learned space were simply these characteristics, their tabulated values would do as well as the learned matrix.
They do not.
A fixed matrix made of a constant column and the standardized Pauling electronegativity, logarithm of the molar volume, and d-shell occupancy (\fchr) gives 177~\mev\ in the linear model and 102~\mev\ in the network model for 86 elements at $\Npe = 4$ and $N = 10^5$, against 159 and 85 for \fsubp and 84 and 46 learned (Table~\ref{tab:ee}).
For the 25 metals, it gives 82~\mev\ in the linear model ($N = 10^5$) and 49~\mev\ in the network model ($N = 10^4$), against 64 and 48 for \fsubp and 42 and 33 learned.
Miedema's solid-state parameters do better on the metals than the Pauling and elemental scales.
With a constant column, $\phi^*$, $n_\mathrm{ws}^{1/3}$, and $V^{2/3}$ (\fmie) give 62~\mev\ (linear) and 44~\mev\ (network), as good as \fsubp and the best published embeddings, and closing half of the gap to the learned space in the linear model and a third in the network model.
Using the d occupancy in place of $V^{2/3}$ does not help (64 and 48~\mev).
The tabulated characteristics are thus about as useful as substitution statistics, somewhat worse than the best published embeddings, and give about twice the error of the learned space.
They span a large part of it: their subspace overlap with the network model's four-dimensional space is 0.63 for 86 elements and 0.80 for the metals, against 0.25 and 0.35 when random columns replace the three characteristics.
The shortfall lies mostly in one direction.
For 86 elements, the cosines of the principal angles between the two spaces are 0.98, 0.92, 0.80, and 0.25 (1.00, 0.97, 0.95, and 0.61 for the metals): three learned directions lie close to the span of the characteristics, and the fourth is nearly missing from it.
This direction separates the alkali metals and the elements with a filled or nearly filled d shell (Cu, Ag, Au, Pd, Cd, Hg) from Ce, the early actinides, and the heavy p-block elements (Sb, Bi, Te, I).
It does not single out the light nonmetals, whose tabulated molar volumes are those of molecular solids.
Nonlinear distortions of the characteristics recover part of the rest: quadratic functions of them raise the leave-one-out $R^2$ of the learned coordinates from 0.47 to 0.72, and cubic ones overfit.
The learned space thus contains the tabulated scales, somewhat adjusted, together with one direction that they lack.

Recovering Miedema's parameters from DFT formation energies alone is consistent with the success of his model, in which two or three numbers per element account for most of the formation enthalpy of metallic alloys.
The extra dimensions needed across 86 elements may carry what his model handles with additional hybridization terms, and the ionic and covalent bonding that lies outside its scope.

\section{Discussion}

The two routes rest on different information, structures in one case and compositions in the other.
They agree that about four learned dimensions capture most of what a single property needs to know about chemical identity, and that six to eight suffice for several properties at once.
Ten times more training structures lower every error but leave this picture unchanged, so the dimension reflects the chemistry rather than the amount of data.
This extends the two-to-four pseudoelements found for restricted chemical spaces\cite{willatt2018,lopanitsyna2023} to 86 elements.
Two of these dimensions have a clear physical meaning, an electronegativity scale and an atomic size, which are the ingredients of Pettifor's chemical scale and of Miedema's model\cite{pettifor1984,miedema1980}.
A third follows the filling of the d shell, which the seeds and the linear model reproduce.
These are the atomic quantities of tight-binding descriptions of cohesion\cite{pettifor1995}: the energies of the valence orbitals, which drive charge transfer, their radii, which set the hopping and the overlap repulsion, and the filling of the d band, which sets the balance of bonding and antibonding states.
The learned four-dimensional spaces recover the LDA eigenvalues of the valence s and p levels of the free atoms\cite{kotochigova1997} with leave-one-out $R^2 = 0.69$--0.84 and their calculated radii\cite{clementi1967} with 0.66--0.82, about as well as the tabulated electronegativity and atomic radius (Table~S6).
They recover the energy of the d level much less well than the number of d electrons (Section~\ref{sec:meaning}): the third direction follows the filling of the d band rather than the position of the d level.
The irregularities discussed below also have known shell origins: the poor screening by the 4f electrons behind the lanthanide contraction, the half-filled and filled 4f shells that make Eu and Yb divalent, and the near-degeneracy of the 4f and 5d levels of Ce, consistent with its place near Th.
Beyond these continuous characteristics, the extra dimensions mostly keep the columns of the periodic table together.
The learned space is therefore best read as a small set of element characteristics, with neighbor densities weighted by them, rather than as a set of pseudoelements.

The periodic table already arranges the elements by similarity, in two dimensions, so it is fair to ask what the learned space adds.
The table gets the topology right: in the learned space, 46\% of each element's nearest neighbors share its column, and the position in the table has the highest family purity of any space we tested (0.77).
What it lacks is a metric.
Its axes are not the directions that energies use: electronegativity and size each run diagonally across the table, which is why diagonal pairs such as Be--Al and B--Si come out as close neighbors.
Its steps all have the same length, whereas the chemical steps do not: the 4d and 5d metals of a group are closer than the 3d and 4d ones, Eu and Yb stand apart from the other lanthanides, and Sc lies closer to Lu than to La.
And it is two-dimensional only as a drawing: written as coordinates, with the f block cut out and the d block inserted, it has rank six, and it still lacks the d filling as a continuous scale that the f block continues, as well as the fourth learned direction.
As a fixed space, the position in the table recovers 59\% of the gain of the learned space at $\Npe = 4$, and all seven of its dimensions give 143~\mev, against 64~\mev\ for eight learned ones.
Learned only as a smooth function of the position in the table, the element space of the network model comes within 15--30\% of an unrestricted one (Section~\ref{sec:meaning}): the table's coordinates carry most of the information, in distorted form, and what they miss matches these irregularities.

The practical consequence is that an element space is worth learning once and reusing.
A four-dimensional matrix learned on $10^5$ PBE structures and kept fixed beats chemical-knowledge matrices and the best published embeddings on small data sets, and carries over to the energy above the hull, to the magnetization, and to PBEsol and SCAN formation energies.
Within one functional, a network model gains little by adapting it further; across functionals, it gains clearly, and does best when it starts from the transferred matrix rather than from substitution statistics.
Published embeddings are no substitute: fifteen of them, including those of universal interatomic potentials, are no better than substitution statistics as low-dimensional bottlenecks.
The fraction of the learned space contained in an embedding's leading components predicts how well it will compress, and gives a simple test for new embeddings.

For a model that has to cover many elements, the results translate into concrete choices.
For 86 elements, channels beyond 8--12 no longer lower the formation-energy error, even with $10^6$ training structures, and a contraction to 8 or 12 channels shrinks the power spectrum from about $1.2 \times 10^6$ to $1.0 \times 10^4$ or $2.3 \times 10^4$ features per atom, a factor of 50 to 100.
Four channels save another factor of four, at the price of an error about a quarter higher (23 against 18~\mev\ at $N = 10^6$ with 20 epochs).
The same saving applies to any descriptor whose cost grows with the square of the number of element channels\cite{bartok2013,musil2021}.
We have not tested the learned spaces inside an interatomic potential, where forces and higher body orders may need more or different channels; the matrices are provided so that this can be tried directly (Data availability).

Two technical points matter for anyone who measures or uses such a dimension.
First, a fixed contraction should keep an element-blind channel: centered components discard the total density, and restoring it lowers the error of \fsub by 5--15\% in the linear model and by 9--19\% in the network model for the metals, although for 86 elements a network model does as well or up to 7\% better without it.
Second, a flexible model can make the dimension meaningless, either through a readout that recovers which elements are present or through a central atom that bypasses the contraction.
Fixed-degree readouts and explicit control of the central atom avoid both.

The study has limitations.
In the structural route, matrices are learned on formation energy and magnetization only; a band-gap transfer test was inconclusive, because the descriptor explains little of the gap (about 0.22~eV against 0.25~eV for the composition baseline).
Three seeds were run at the key points, with standard deviations of at most 1.3~\mev\ at $N = 10^5$, and two at $N = 10^6$, which agree to 0.1~\mev; the rest of the dimension sweep uses one seed.
At $N = 10^6$, the full sweeps use 10 epochs; 20 epochs lower the errors by 8--24\% at $\Npe = 2$--32 for both matrices, the most at small $\Npe$, without changing the optimum.
All data come from Alexandria; the transfer across functionals was tested on Alexandria's PBEsol and SCAN calculations, not on another database.
The descriptor settings (6~\AA\ cutoff, $l \le 4$, 8 radial functions) are fixed throughout.
Published embeddings are compressed by principal components, which depend on feature scaling; a reduction that uses the target could do better, but it would no longer be a fixed baseline.
Finally, the composition models are deliberately simple, and only how their error changes with $\Npe$ is used.

\section{Methods}

\subsection{Data and splits}

\textbf{Structures.}
We use the PBE data of Alexandria (release 2025.07.02)\cite{schmidt2023}, restricted to structures with at most 40 sites and $|E_\mathrm{form}| \le 5$~eV/atom.
We consider two element sets.
The first contains the 25 d-block metals studied by Lopanitsyna et al.\cite{lopanitsyna2023} (Sc--Zn, Y, Zr, Nb, Mo, Ru, Rh, Pd, Ag, Lu, Hf, Ta, W, Ir, Pt, Au).
The second contains 86 common elements.
Each chemical system (the set of elements in a compound) is assigned to a split by a hash of its sorted element symbols: 10\% to test, 5\% to validation, and the rest to training.
As a result, no test compound shares its element set with a training compound.
The test sets contain 23,241 structures (metals) and 30,000 structures (86 elements), and the validation sets contain 10,000 each.
Training sets of $N = 10^3$, $10^4$, and $10^5$ structures are drawn from a pool of 120,000 training structures, and the $N = 10^6$ runs use a separate draw of $10^6$ structures from the 4.8 million available for 86 elements.
All but two of the 86 elements appear in at least 232 of the 120,000 pooled training structures (Yb being the rarest); Xe appears in four and Kr in none (in the $10^6$ draw, Xe in ten and Kr in four), so their vectors in the learned structural spaces are not determined by the data.
We leave them out of Figs.~\ref{fig:maps}c and~\ref{fig:physics}, of Figs.~S1--S4, and of every comparison involving a learned structural space for 86 elements: subspace overlaps, captured fractions, distance and nearest-neighbor statistics, and fits of elemental properties, which are all computed over the other 84 elements.
Family purities are computed over all 86 elements.
Besides the formation energy, we use the energy above the convex hull and the magnetization per atom, $|M|/N_\mathrm{sites}$ with $M$ the total magnetic moment of the cell; about 30\% of the structures have more than 0.05~$\mu_\mathrm{B}$/atom.
\textbf{Other functionals.} Alexandria also provides PBEsol and SCAN calculations for 415,419 compounds. We apply the same filters and the same chemical-system hash, which gives 347,772 training structures for 86 elements (120,000 used) and 21,095 for the 25 metals, with test sets of 30,000 and 2,593 structures.
As an external check, we also evaluate on Materials Project structures\cite{jain2013} in the same element sets, after removing the mean offset between the two databases.

\textbf{Compositions.}
For the composition route, we keep the lowest-energy polymorph of each reduced composition in Alexandria with at most eight elements.
This gives 3,258,859 compositions, split by chemical system in the same way (2,769,259 for training, 165,218 for validation, and 324,382 for testing).
The targets are formation energy, PBE band gap, and volume per atom.

\subsection{Structural route}

For each atom $i$, the neighbor density within 6~\AA\ is expanded in 8 radial functions and spherical harmonics up to $l = 4$, separately for each neighbor element $a = 1, \ldots, \Nel$, giving coefficients $c^i_{a n l m}$, where $n$ indexes the radial functions and $l, m$ the spherical harmonics.
The contraction
\begin{equation}
  \tilde c^{\,i}_{\alpha n l m} = \sum_{a=1}^{\Nel} U_{a\alpha}\, c^i_{a n l m}, \qquad \alpha = 1,\ldots,\Npe,
\end{equation}
with an $\Nel \times \Npe$ matrix $U$, maps the $\Nel$ elements onto $\Npe$ pseudoelements.
The invariant power spectrum is then built from $\tilde c$, which gives $(l_\mathrm{max}+1)\,B(B+1)/2$ features per atom, with $B = 8\Npe$.
The model depends only on the column space of $U$, so $U$ is kept orthonormal.
Every model also includes a composition term with one weight per element.

\textbf{Linear model (\lin).}
Formation energy is a ridge regression on the power spectrum averaged over each structure.
The matrix $U$ is learned by variable projection: at each L-BFGS step, the ridge weights are solved exactly for the current $U$, and the gradient flows only through $U$.
We select the regularization on the validation set from $10^{-8}$ to $10^{2}$.

\textbf{Network model (\net).}
A per-atom neural network (three hidden layers of 1024, 512, and 256 units) maps the power spectrum to atomic energies.
The atomic energies are averaged over each structure and fitted to the residual of the composition term.
$U$ is trained jointly with the network model.
The nonlinearity acts only after the linear contraction, so the neighbors' chemistry still passes through $\Npe$ channels.
We train with AdamW, a one-cycle learning-rate schedule, a smooth-$L_1$ loss, and batches of 256 structures, for 300, 60, 30, and 10 epochs at $N = 10^3$, $10^4$, $10^5$, and $10^6$ (and 20 epochs for a convergence check at $N = 10^6$), and keep the epoch with the lowest validation error.
Unless stated otherwise, $U$ starts from \fsubp.
To test the periodic table (Section~\ref{sec:meaning}), $U = BW$ is learned within the span of a fixed basis $B$, with only $W$ trained and started from the projection of \fsubp onto that span.
$B$ is an orthonormal basis of the polynomials of degree at most 1, 2, or 3 in the standardized coordinates of \fpt (period, group, position in the f row, and block), of rank 7, 17, and 28; the controls are random bases of the same rank that include a constant column ($N = 10^5$, 30 epochs, one seed).

\textbf{Central-atom leak.}
If the density includes the central atom itself, its contribution sits at $r = 0$ and falls almost entirely in the first radial channel, where it is not mixed with neighbors.
The nonlinear network model can then read the central atom's identity from those $\Npe$ numbers, so $\Npe$ bounds only the neighbor chemistry.
We quantify this with two controls.
In the first, the central atom is removed from the density.
In the second, it is removed from the density and given to the network model as a separate one-hot vector, which reintroduces the central atom's identity in full.
The linear model is not affected, because it can only use the self term linearly.

\subsection{Composition route}

Each element $a$ is given a vector $z_a \in \mathbb{R}^{\Npe}$, and a composition with fractions $x_a$ is mapped to $\mu = \sum_{a=1}^{\Nel} x_a z_a$.
Each property is modeled as a polynomial of degree $d$ in the components $\mu_\alpha$ ($\alpha = 1, \ldots, \Npe$), plus a linear term in the fractions.
For $d = 3$, the highest degree used here,
\begin{equation}
  y = \sum_{a=1}^{\Nel} w_a x_a
    + \sum_{\alpha \le \beta} v_{\alpha\beta}\, \mu_\alpha \mu_\beta
    + \sum_{\alpha \le \beta \le \gamma} v_{\alpha\beta\gamma}\, \mu_\alpha \mu_\beta \mu_\gamma ,
\end{equation}
and the $d = 2$ model keeps only the first two sums.
Higher degrees add sums over four or more pseudoelement indices in the same way.
The quadratic term equals $\sum_{ab} x_a x_b\, z_a^\top V z_b$ with a symmetric $\Npe \times \Npe$ matrix $V$, a pair interaction between elements of rank at most $\Npe$.
The cubic term is a three-body interaction of the same structure.
The degree is fixed deliberately.
A flexible readout such as a neural network could recover which elements are present from $\mu$ alone, even when $\Npe = 1$, and $\Npe$ would then measure nothing.
With \fid ($Z = I$, $\Npe = \Nel$), the $d = 2$ model contains every pair interaction and serves as the uncompressed reference.
$Z$ is learned by the same variable projection on 400,000 training compositions, either for all three properties jointly or for each property alone.
The final ridge weights are then refit on the full training set.

\subsection{Fixed element spaces}

We compare learned spaces with the following fixed ones.
\begin{itemize}
  \item \textbf{\fsub.} The top-$\Npe$ principal components of the element substitution matrix (98 dimensions) of Wang et al.\cite{wang2021}, mined from the Inorganic Crystal Structure Database, normalized as in Cerqueira et al.\cite{cerqueira2026}.
  \item \textbf{\fsubp.} The top $\Npe-1$ components of \fsub plus a constant column. The constant column carries the element-blind total density, which centered components otherwise discard.
  \item \textbf{\fmv.} The top-$\Npe$ principal components of the MatterVial Global element embeddings (832 dimensions)\cite{mattervial}.\footnote{In the published MatterVial dictionary, the two 64-dimensional blocks attributed to the ROOST models (band gap and OQMD formation energy) are identical for every element, although the two models have different weights. The dictionary therefore has 768 distinct dimensions, and the ROOST features count twice. We used the dictionary as distributed. Removing one copy barely changes the embedding: the leading 2--16 principal subspaces overlap with the original ones at 0.97--1.00, and the pairwise element distances keep a rank correlation of 0.99.}
  \item \textbf{Published embeddings.} The fifteen element embeddings distributed with ElementEmbeddings 0.8.0\cite{onwuli2023}, used like \fsubp (top $\Npe-1$ principal components plus a constant column). They are Magpie\cite{ward2016}, Oliynyk\cite{oliynyk2016}, and XenonPy\cite{kusaba2022,kusaba2023} (tabulated properties); mat2vec\cite{tshitoyan2019}, MatScholar\cite{weston2019}, and MatSciBERT\cite{gupta2022} (literature); MEGNet\cite{chen2019}, SkipAtom\cite{antunes2022}, CGNF\cite{jang2024}, CrystaLLM\cite{antunes2024}, and Chemeleon\cite{park2025} (models of crystal structures); and CHGNet\cite{deng2023}, MACE-MP-0\cite{batatia2024}, ORB v2\cite{neumann2024}, and SevenNet\cite{park2024} (universal interatomic potentials).
  \item \textbf{\fmvp.} The top $\Npe-1$ components of \fmv plus a constant column.
  \item \textbf{\fpt.} The period, the group (3 for the f block), the position in the f row (0 outside it), and a one-hot block, standardized, treated like the published embeddings.
  \item \textbf{\fchr.} A constant column and the standardized Pauling electronegativity, logarithm of the molar volume, and $(n-1)$d-shell occupancy (from the ground-state electron configuration, zero for elements without d electrons below the valence shell), all from pymatgen\cite{ong2013}. The composition model uses the three characteristics without the constant column.
  \item \textbf{\fmie.} A constant column and Miedema's $\phi^*$, $n_\mathrm{ws}^{1/3}$, and $V^{2/3}$\cite{miedema1980}, standardized, for the 25 metals only.
  \item \textbf{\frnd.} A random orthonormal matrix.
  \item \textbf{\fid.} The identity matrix, $\Npe = \Nel$, which leaves the element channels uncompressed.
  \item \textbf{\flrn.} The matrix \lrn learned at $\Npe = 4$ (or 8) on $10^5$ PBE structures for the formation energy, kept fixed in a new task.
\end{itemize}
Principal components depend on how features are scaled.
We standardize each feature of the three property-based embeddings, whose features have different units, and use the other embeddings as distributed.
The Oliynyk table lacks four actinides (Ac, Pa, Np, Pu), which we set to the feature means.
This choice matters for some embeddings: standardizing every feature lowers SkipAtom's family purity (Section~\ref{sec:measure}) from 0.71 to 0.22, and raw ORB v2 vectors are dominated by a single direction, which alone holds 90\% of their variance.
In the composition route, we weight the principal components by how often each element combination occurs.
For an embedding matrix $E$ and the covariance $\mathrm{Cov}(x)$ of training-set fractions, we take the top eigenvectors of $E^\top \mathrm{Cov}(x)\, E$, which form the latent space of a linear autoencoder of the compositions' vectors $\sum_a x_a e_a$.
A learned linear map applied to MatterVial vectors would not be a fixed baseline, because the $86 \times 832$ matrix has full rank and such a map could produce any embedding.

\subsection{Comparing element spaces}

To compare two learned spaces, we use three measures:
\begin{itemize}
  \item subspace overlap, the mean squared cosine of the principal angles between the two column spaces (chance level $\Npe/\Nel$);
  \item the Spearman correlation between their pairwise element distances;
  \item the fraction of shared five nearest neighbors.
\end{itemize}
To ask how much of a learned space a fixed embedding contains, we take the mean squared canonical correlation between the embedding's top-$p$ principal components and the learned space.
This captured fraction is 1 if the learned space lies within those $p$ components, and about $p/\Nel$ for random ones.
The family purity (Section~\ref{sec:measure}) is computed on the orthonormalized learned matrices and on the embeddings scaled as above; its chance level, 0.20, is the mean over elements of the fraction of the other elements that share the element's family.
Leaving out Kr and Xe raises every purity by 0.01--0.03 and does not change which spaces score highest.
To interpret a learned space, we fit elemental properties linearly with an intercept and report leave-one-out $R^2$, and we find the directions shared with eight textbook properties by canonical correlation analysis.
As free-atom quantities, we use the LDA eigenvalues of the NIST reference data\cite{kotochigova1997} (through pymatgen) for the outermost s level, the outermost p level of the p-block elements, the outermost d level of the d-block elements, and the highest occupied level, leaving out the s and p levels of Pd, which lie below its filled 4d shell, and the radii calculated by Clementi et al.\ from Hartree--Fock orbitals\cite{clementi1967}; the eigenvalues are not available for Np and Pu, nor the radii for La, Ce, Ho, and the actinides.
To remove two of these directions, we project the centered element coordinates onto the orthogonal complement of their coefficient vectors.
The d-filling direction is then the linear fit of the d-electron count (group number minus 2, for the d-block metals of groups 3--12) within that complement.
For the periodic-table analyses, the column of an element is its group, with Sc, Y, and Lu in group 3 and La--Yb and Ac--Pu forming a separate f column (for the family purity, Lu counts instead among the f-block metals); diagonal neighbors differ by one period and one group.
To ask whether the channels behave as pseudoelements, we rotate the orthonormalized $U$ by varimax, which makes each column as localized on few elements as possible, and measure the number of elements carrying each column, $1/\sum_a U_{a\alpha}^4$, and the weight $\sum_a U_{a\alpha}^2$ of its entries of minority sign.

\subsection{Use of AI tools}
Claude models (Anthropic), used through the Claude Code assistant, and Gemini (Google), used through Google Antigravity, helped write and run the training and analysis scripts, draft and edit the text, review the drafts of the manuscript and suggest corrections, and check the references against their DOI records. The author designed and directed the study, checked the results and the text, and takes full responsibility for the content.

\backmatter

\bmhead{Data availability}
The Alexandria and Materials Project data are publicly available\cite{schmidt2023,jain2013}.
The learned element spaces of both routes, for every dimension studied, the identifiers of all training, validation, and test structures, and the composition table of the composition route are available in the Materials Cloud Archive\cite{talirz2020} (\href{https://doi.org/10.24435/materialscloud:jg-04}{doi:10.24435/materialscloud:jg-04}).

\bmhead{Code availability}
The training and analysis scripts, with an example of how to load and apply a learned element space, are available in the same Materials Cloud Archive record (\href{https://doi.org/10.24435/materialscloud:jg-04}{doi:10.24435/materialscloud:jg-04}).

\bmhead{Acknowledgements}
This project has received funding from the European Research Council
(ERC) under the European Union’s Horizon Europe research and innovation
programme (Grant agreement No. 101224255 - GEMPROMISE).
G.-M.R. also acknowledges financial support from the Fonds de la Recherche Scientifique de Belgique (FRS-FNRS). Computational resources have been provided by the Consortium des \'Equipements de Calcul Intensif, funded by the FRS-FNRS under Grant No.~2.5020.11 and by the Walloon Region.

\bmhead{Author contributions}
G.-M.R. conceived and directed the study, performed the calculations and analysis, and wrote the manuscript, with the assistance of AI tools as described in Methods.

\bmhead{Competing interests}
The author declares no competing interests.


\end{document}


\maketitle

Tables~S1--S4 give the central-atom controls of the network model, the task dependence of the composition route, the agreement between learned spaces, and the reconstruction dimension of existing embeddings.
Table~S5 gives the dimension sweep for the magnetization, Table~S6 the elemental properties recovered from the learned four-dimensional spaces, Table~S7 the formation-energy error by compound class, Table~S8 compares the transferred element space with published embeddings, and Table~S9 gives the warm start of the network model from the transferred matrix on other functionals.
Figs.~S1 and~S2 show the learned structural spaces on the periodic table and after removing the electronegativity and size directions, and Figs.~S3 and~S4 follow the four directions of Fig.~S1 down the periods and along the columns of the table.
Notation and methods follow the main text.

\begin{table}[h!]
\centering
\small
\caption{Central-atom controls for the network model: formation-energy test error (\mev) at $N = 10^4$ with a learned matrix, one seed per entry. Differences of a few \mev\ between entries, such as the rise from 79 to 83~\mev\ for 86 elements with a separate one-hot input, are comparable to the standard deviation between seeds at this training-set size (up to 4~\mev, Table~3 of the main text).}
\label{tab:center}
\begin{tabular}{llccc}
\toprule
Elements & $\Npe$ & Central atom in density & Removed & Removed, one-hot input \\
\midrule
25 metals & 4 & 33 & 39 & 34 \\
25 metals & 8 & 34 & 39 & 37 \\
86 elements & 4 & 75 & 106 & 79 \\
86 elements & 8 & 70 & 98 & 83 \\
\bottomrule
\end{tabular}
\end{table}

\begin{table}[h!]
\centering
\small
\caption{Composition models of degree $d = 3$ (\plthree) for 86 elements: test error for each property, with element vectors learned for that property alone or for all three jointly.}
\label{tab:single}
\begin{tabular}{llrrrrr}
\toprule
Property & Learned on & $\Npe = 2$ & 3 & 4 & 6 & 8 \\
\midrule
Formation energy (\mev) & itself & 229 & 209 & 206 & 191 & 184 \\
 & all three & 278 & 235 & 220 & 205 & 194 \\
Band gap (eV) & itself & 0.253 & 0.250 & 0.251 & 0.240 & 0.237 \\
 & all three & 0.306 & 0.267 & 0.261 & 0.248 & 0.244 \\
Volume (\AA$^3$/atom) & itself & 1.65 & 1.61 & 1.57 & 1.52 & 1.49 \\
 & all three & 1.79 & 1.78 & 1.72 & 1.62 & 1.56 \\
\bottomrule
\end{tabular}
\end{table}

\begin{table}[h!]
\centering
\small
\caption{Agreement between learned spaces at $\Npe = 4$: subspace overlap (chance level in parentheses), Spearman correlation of pairwise element distances, and fraction of shared five nearest neighbors. Linear matrices are fit on 32,000 structures and the network model on 100,000. For 86 elements, the comparisons are made over 84 elements, without Kr and Xe, whose learned positions are not set by the training data.}
\label{tab:space}
\begin{tabular}{llccc}
\toprule
Elements & Comparison & Overlap & Distance Spearman & Shared 5-NN \\
\midrule
25 metals & \lin, two random starts & 0.74 (0.16) & 0.68 & 63\% \\
25 metals & \lin vs.\ \net & 0.84 (0.16) & 0.82 & 76\% \\
25 metals & \net vs.\ its starting matrix & 0.88 (0.16) & 0.72 & 74\% \\
\midrule
86 elements & \lin, two random starts & 0.71 (0.05) & 0.83 & 48\% \\
86 elements & \lin vs.\ \net & 0.65 (0.05) & 0.76 & 56\% \\
86 elements & \net vs.\ its starting matrix & 0.81 (0.05) & 0.75 & 49\% \\
\bottomrule
\end{tabular}
\end{table}

\begin{table}[h!]
\centering
\small
\caption{Principal components needed for 90\% and 80\% of the variance, and the share of variance in the top three components, for element vectors weighted equally or weighted by composition (3.26 million Alexandria compositions).}
\label{tab:pca}
\begin{tabular}{llcc}
\toprule
Embedding & Elements & Equal weights & Composition-weighted \\
 & & 90\% / 80\% / top 3 & 90\% / 80\% / top 3 \\
\midrule
MatterVial Global (832-D) & 86 & 43 / 29 / 27\% & 36 / 22 / 34\% \\
Substitution embedding (98-D) & 86 & 37 / 23 / 37\% & 29 / 17 / 43\% \\
One-hot & 86 & 77 / 68 / 4\% & 61 / 49 / 15\% \\
\midrule
MatterVial Global & 25 metals & 15 & 14 \\
Substitution embedding & 25 metals & 14 & 14 \\
One-hot & 25 metals & 22 & 20 \\
\bottomrule
\end{tabular}
\end{table}

\begin{table}[h!]
\centering
\small
\caption{Magnetization per atom: network test error (m$\mu_\mathrm{B}$/atom) at $N = 10^5$ against the dimension of the element space, with a matrix learned on the magnetization (\lrn) or fixed to \fsubp{} (seed 0, 30 epochs). Baseline: the composition baseline of the main text, without any element space ($\Npe = 0$).}
\label{tab:mag}
\begin{tabular}{lrrrrrrrrr}
\toprule
 & \multicolumn{6}{c}{\lrn, $\Npe =$} & \multicolumn{2}{c}{\fsubp{}} & \\
\cmidrule(lr){2-7}\cmidrule(lr){8-9}
Elements & 2 & 3 & 4 & 6 & 8 & 12 & 4 & 8 & Baseline \\
\midrule
25 metals & 57 & 45 & 41 & 41 & 43 & 46 & 50 & 46 & 214 \\
86 elements & 59 & 53 & 51 & 49 & 50 & 51 & 63 & 58 & 138 \\
\bottomrule
\end{tabular}
\end{table}

\begin{table}[h!]
\centering
\small
\caption{Leave-one-out $R^2$ of elemental properties fitted linearly from the learned four-dimensional spaces of the three models (models for 86 elements, fitted over 84 of them, without Kr and Xe; formation energy for the structural models, three properties jointly for the composition model). Miedema's parameters are tabulated for 73 of the 86 elements, mostly metals. Valence electrons count the s and p electrons of main-group elements and the group number up to group 12. The last five rows are free-atom quantities: LDA eigenvalues of the outermost s, p (p block), and d (d block) levels and of the highest occupied level, from the NIST reference data, and the radii calculated from Hartree--Fock orbitals by Clementi et al.\ (Methods of the main text). The random column is lower for fewer elements.}
\label{tab:physics}
\begin{tabular}{lcrrrr}
\toprule
Property & Elements & \net & \lin & \plthree & Random \\
\midrule
Miedema $\phi^*$ & 73 & 0.89 & 0.86 & 0.94 & $-0.09$ \\
Miedema $n_\mathrm{ws}^{1/3}$ & 73 & 0.85 & 0.77 & 0.86 & $-0.09$ \\
Miedema $V^{2/3}$ & 73 & 0.72 & 0.51 & 0.82 & $-0.09$ \\
\midrule
Pauling electronegativity & 84 & 0.77 & 0.78 & 0.84 & $-0.08$ \\
Mendeleev number & 84 & 0.80 & 0.76 & 0.81 & $-0.08$ \\
First ionization energy & 84 & 0.64 & 0.62 & 0.68 & $-0.08$ \\
Atomic radius & 84 & 0.71 & 0.60 & 0.77 & $-0.08$ \\
Molar volume & 84 & 0.65 & 0.47 & 0.58 & $-0.08$ \\
Group & 84 & 0.82 & 0.75 & 0.78 & $-0.08$ \\
d electrons (d block) & 30 & 0.85 & 0.81 & 0.82 & $-0.25$ \\
Valence electrons & 84 & 0.43 & 0.39 & 0.47 & $-0.08$ \\
Period & 84 & 0.40 & 0.39 & 0.48 & $-0.08$ \\
Electron affinity & 84 & 0.41 & 0.38 & 0.40 & $-0.08$ \\
\midrule
Valence s level & 81 & 0.74 & 0.69 & 0.77 & $-0.08$ \\
Valence p level (p block) & 23 & 0.76 & 0.80 & 0.84 & $-0.35$ \\
Valence d level (d block) & 30 & 0.36 & 0.36 & 0.53 & $-0.25$ \\
Highest occupied level & 82 & 0.51 & 0.52 & 0.59 & $-0.08$ \\
Orbital radius (Clementi) & 75 & 0.71 & 0.66 & 0.82 & $-0.09$ \\
\bottomrule
\end{tabular}
\end{table}

\begin{table}[h!]
\centering
\small
\caption{Formation-energy test error (\mev) by compound class, 86 elements, $N = 10^5$, seed 0. Metallic: metals only. Fixed valence: nonmetals or metalloids with metals of a single common oxidation state only (groups 1--3, Zn, Cd, Al, Ga, In, and the trivalent lanthanides). Multivalent: nonmetals or metalloids with at least one other metal (d metals of groups 4--11, Sn, Pb, Tl, Bi, Ce, Sm, Eu, Yb, and the actinides). Nonmetal: nonmetals and metalloids only. Metals are those of pymatgen. The number of test structures is given in parentheses.}
\label{tab:strata}
\begin{tabular}{llrrrrr}
\toprule
Model & $U$ & All & Metallic & Fixed valence & Multivalent & Nonmetal \\
 & & (30,000) & (13,681) & (3,199) & (12,895) & (225) \\
\midrule
\net & \lrn, $\Npe = 2$ & 98.6 & 78.1 & 111.2 & 116.1 & 157.8 \\
 & \lrn, $\Npe = 3$ & 56.8 & 42.4 & 68.6 & 68.1 & 114.9 \\
 & \lrn, $\Npe = 4$ & 45.7 & 33.8 & 56.4 & 54.9 & 86.9 \\
 & \lrn, $\Npe = 6$ & 38.3 & 27.7 & 48.9 & 46.4 & 73.8 \\
 & \lrn, $\Npe = 8$ & 35.6 & 25.2 & 46.4 & 43.3 & 71.0 \\
 & \lrn, $\Npe = 12$ & 35.8 & 24.8 & 46.0 & 44.1 & 76.1 \\
 & \lrn, $\Npe = 16$ & 36.6 & 25.0 & 48.8 & 45.0 & 85.9 \\
 & \fsubp{}, $\Npe = 4$ & 85.5 & 67.8 & 101.9 & 99.2 & 141.4 \\
 & \fchr, $\Npe = 4$ & 102.2 & 82.2 & 129.0 & 114.9 & 210.1 \\
\midrule
\lin & \lrn, $\Npe = 3$ & 99.2 & 69.0 & 135.1 & 120.4 & 212.4 \\
 & \lrn, $\Npe = 4$ & 84.5 & 58.5 & 120.9 & 101.0 & 196.9 \\
 & \lrn, $\Npe = 6$ & 71.0 & 47.8 & 101.0 & 86.6 & 164.2 \\
 & \lrn, $\Npe = 8$ & 64.5 & 42.7 & 95.4 & 78.2 & 161.4 \\
 & \lrn, $\Npe = 12$ & 65.0 & 38.7 & 101.7 & 81.1 & 216.7 \\
 & \lrn, $\Npe = 16$ & 65.8 & 42.1 & 100.0 & 80.5 & 177.1 \\
 & \fsubp{}, $\Npe = 4$ & 159.4 & 118.9 & 192.0 & 190.4 & 385.0 \\
\bottomrule
\end{tabular}
\end{table}

\begin{table}[h!]
\centering
\footnotesize
\caption{Reusing a four-dimensional element space: the transferred matrix (\flrn, as in Table~3 of the main text) against three published embeddings, XenonPy, CGNF, and MACE-MP-0 (MACE), kept fixed at $\Npe = 4$ (top three principal components plus a constant column, as in Table~2 of the main text), for $N = 10^3$ and $10^4$ training structures. Errors are in \mev, except for the magnetization (m$\mu_\mathrm{B}$/atom); mean over three seeds. The standard deviation between seeds is at most 14 at $N = 10^3$ and at most 3 at $N = 10^4$. The best entry of each row and training-set size is in bold.}
\label{tab:xfer}
\setlength{\tabcolsep}{3pt}
\begin{tabular}{lllrrrrrrrr}
\toprule
 & & & \multicolumn{4}{c}{$N = 10^3$} & \multicolumn{4}{c}{$N = 10^4$} \\
\cmidrule(lr){4-7}\cmidrule(lr){8-11}
Target & Model & Elements & XenonPy & CGNF & MACE & \flrn & XenonPy & CGNF & MACE & \flrn \\
\midrule
PBE $E_\mathrm{form}$ & \lin & 25 metals & 130 & 109 & 122 & \textbf{78} & 96 & 81 & 86 & \textbf{50} \\
 &  & 86 elements & 281 & 243 & 287 & \textbf{178} & 185 & 170 & 197 & \textbf{101} \\
 & \net & 25 metals & 105 & 95 & 101 & \textbf{64} & 50 & 47 & 53 & \textbf{32} \\
 &  & 86 elements & 276 & 268 & 288 & \textbf{195} & 164 & 166 & 171 & \textbf{75} \\
\midrule
PBEsol $E_\mathrm{form}$ & \lin & 25 metals & 136 & 108 & 117 & \textbf{75} & 94 & 76 & 80 & \textbf{43} \\
 &  & 86 elements & 277 & 251 & 294 & \textbf{189} & 185 & 178 & 212 & \textbf{106} \\
 & \net & 25 metals & 121 & 100 & 113 & \textbf{78} & 50 & 46 & 50 & \textbf{35} \\
 &  & 86 elements & 299 & 319 & 342 & \textbf{245} & 176 & 186 & 210 & \textbf{107} \\
\midrule
SCAN $E_\mathrm{form}$ & \lin & 25 metals & 152 & 123 & 136 & \textbf{91} & 113 & 90 & 99 & \textbf{58} \\
 &  & 86 elements & 304 & 280 & 322 & \textbf{205} & 209 & 198 & 237 & \textbf{124} \\
 & \net & 25 metals & 127 & 109 & 121 & \textbf{87} & 58 & 53 & 59 & \textbf{42} \\
 &  & 86 elements & 322 & 340 & 360 & \textbf{267} & 196 & 205 & 230 & \textbf{125} \\
\midrule
$E_\mathrm{hull}$ & \net & 25 metals & 96 & 87 & 95 & \textbf{68} & 54 & 49 & 55 & \textbf{38} \\
 &  & 86 elements & 205 & 212 & 223 & \textbf{175} & 121 & 131 & 137 & \textbf{88} \\
Magnetization & \net & 25 metals & 151 & 134 & 146 & \textbf{116} & 86 & 82 & 85 & \textbf{67} \\
 &  & 86 elements & 151 & 153 & 148 & \textbf{143} & 105 & 102 & 88 & \textbf{79} \\
\bottomrule
\end{tabular}
\end{table}

\begin{table}[h!]
\centering
\small
\caption{Warm start across functionals: network model at $\Npe = 4$ trained on PBEsol or SCAN formation energies with \fsubp kept fixed, with a matrix learned from the \fsubp starting point (\lrn, as in Table~3 of the main text), with the PBE matrix kept fixed (\flrn), and with a matrix learned from the PBE matrix as starting point (warm start). Test error in \mev, mean over three seeds; the standard deviation between seeds is given for the two learned matrices. The best entry of each row is in bold.}
\label{tab:warm}
\begin{tabular}{lllrrrr}
\toprule
Functional & Elements & $N$ & \fsubp & \lrn & \flrn & Warm start \\
\midrule
PBEsol & 86 elements & $10^3$ & 347 & 182 $\pm$ 5 & 245 & \textbf{158} $\pm$ 1 \\
 & & $10^4$ & 210 & 76 $\pm$ 1 & 107 & \textbf{72} $\pm$ 1 \\
 & 25 metals & $10^3$ & 96 & 59 $\pm$ 2 & 78 & \textbf{56} $\pm$ 3 \\
 & & $10^4$ & 47 & 31 $\pm$ ${<}1$ & 35 & \textbf{30} $\pm$ 1 \\
\midrule
SCAN & 86 elements & $10^3$ & 358 & 192 $\pm$ 2 & 267 & \textbf{176} $\pm$ 2 \\
 & & $10^4$ & 226 & 89 $\pm$ 1 & 125 & \textbf{86} $\pm$ 1 \\
 & 25 metals & $10^3$ & 108 & 76 $\pm$ 3 & 87 & \textbf{69} $\pm$ 3 \\
 & & $10^4$ & 56 & 40 $\pm$ 1 & 42 & \textbf{37} $\pm$ 1 \\
\bottomrule
\end{tabular}
\end{table}

\begin{figure}[h!]
\centering
\includegraphics[width=\textwidth]{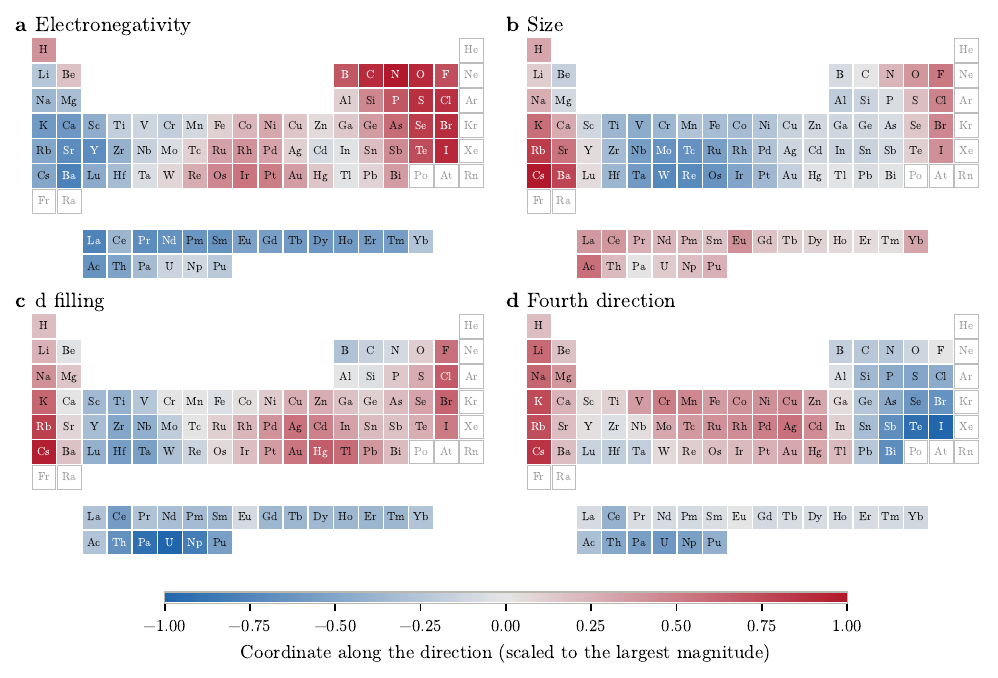}
\caption{The four-dimensional space learned by the network model ($N = 10^5$, 86 elements) drawn on the periodic table, without Kr and Xe, whose positions are not set by the training data.
The directions are fitted in turn: (a) the linear fit of the Pauling electronegativity; (b) the fit of the logarithm of the molar volume, orthogonal to (a); (c) the fit of the d-electron count of the d-block metals, orthogonal to (a) and (b); (d) the remaining direction.
Each cell shows the coordinate of the element along the direction, scaled to the largest magnitude; empty cells are elements outside the 86.
Directions (a) and (b) agree with those of the linear and composition models ($r = 0.82$--0.97); (c) and (d) agree less ($r = 0.14$--0.64).}
\end{figure}

\begin{figure}[h!]
\centering
\includegraphics[width=\textwidth]{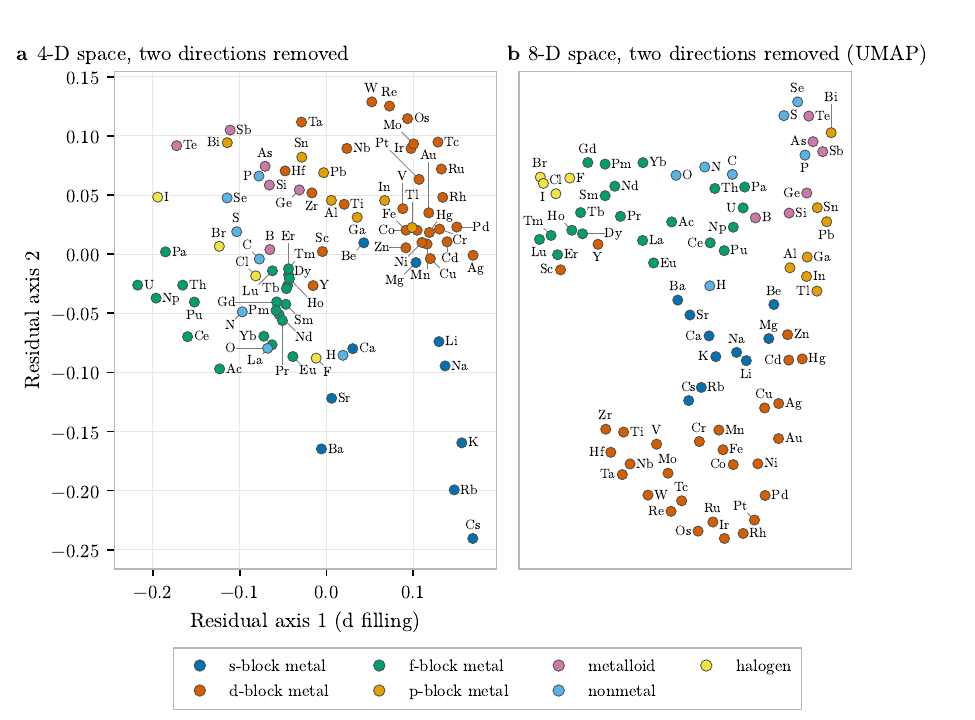}
\caption{Learned structural spaces ($N = 10^5$) after projecting out the electronegativity and size directions of Fig.~5a in the main text, without Kr and Xe.
(a) The two remaining dimensions of the four-dimensional space, with the d-filling axis horizontal.
(b) The six remaining dimensions of the eight-dimensional space, mapped to two dimensions with UMAP (10 neighbors, minimum distance 0.3): the d metals line up by group, and the columns of the periodic table stay together.}
\end{figure}

\begin{figure}[h!]
\centering
\includegraphics[width=0.92\textwidth]{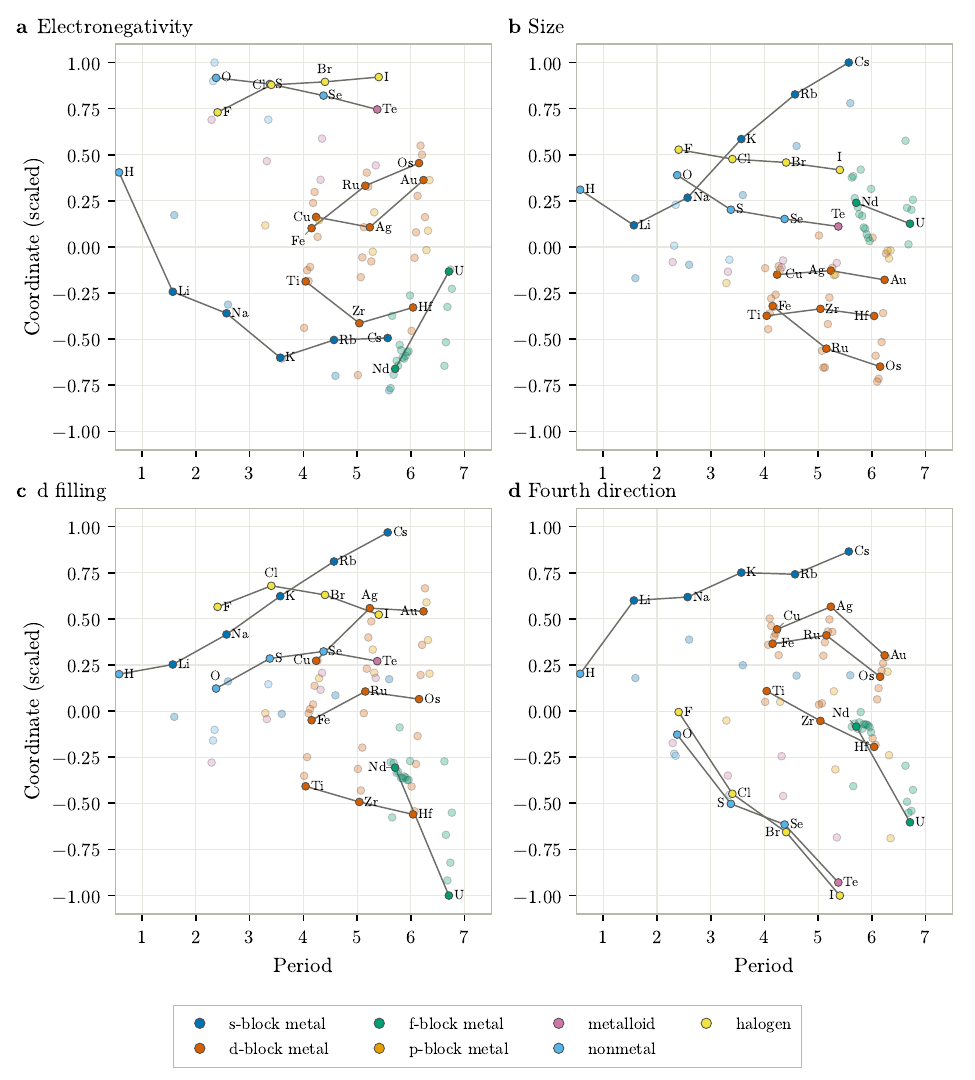}
\caption{The four directions of Fig.~S1 against the period, without Kr and Xe, colored by family as in Fig.~4 of the main text.
Within a period, the elements are spread by their column, from the s block on the left through the f and d blocks to the p block on the right.
Seven columns are joined across periods and labeled: the alkali metals (with H), groups 4, 8, and 11, the chalcogens, the halogens, and Nd and U; the other elements are faded.
Down the d groups, the size coordinate (b) decreases (from Cr to W, for instance), while the molar volume grows from the 3d to the 4d metals.
This comes mostly from fitting the size direction orthogonal to electronegativity, which rises down the late d groups: the unconstrained fit of the logarithm of the molar volume decreases only slightly down these groups, and correlates with it at $r = 0.74$ within the d block, against 0.45 for direction (b).}
\end{figure}

\begin{figure}[h!]
\centering
\includegraphics[width=\textwidth]{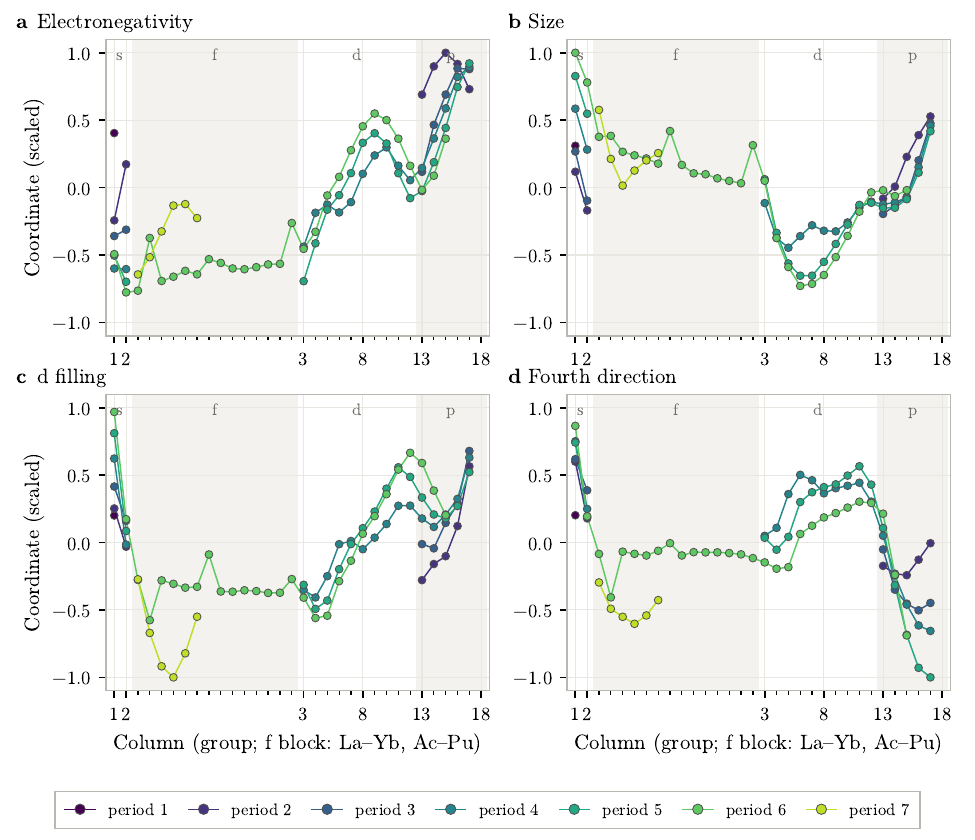}
\caption{The four directions of Fig.~S1 against the column of the 32-column periodic table, without Kr and Xe, colored by period: the s block, the f block (La--Yb and Ac--Pu), the d block (group 3 is Sc, Y, and Lu), and the p block.
Lines join the elements of a period that sit in neighboring columns.
Each direction is ordered along the rows of each block, but the order differs from block to block.
Along the lanthanides, the size coordinate (b) decreases from La to Tm, as the lanthanide contraction does, except at Eu and Yb, which are divalent in many compounds; Ce, which readily takes the 4+ state, departs from the other lanthanides toward Th and the light actinides on directions (a), (c), and (d).}
\end{figure}